\documentclass[12pt]{article}
\usepackage{psfig}
\usepackage{graphics}
\usepackage{color}

\def\sk{\vskip 1em}
\def\noi{\noindent}

\def\al{\alpha}
\def\be{\beta}
\def\ga{\gamma}
\def\th{\Theta}
\def\ep{\epsilon}
\def\l{\ell}

\def\Ah{\widehat{A}}
\def\Dh{\hat{D}}
\def\phih{\widehat{\phi}}

\def\Fh{\widehat{F}}

\def\CLh{\widehat{\CL}}

\def\Ga{\Gamma}
\def\del{\delta}
\def\d{{\mbox{\tiny{$D$}}}}
\def\D{\delta}

\def\la{\lambda}
\def\sma#1{\mbox{\footnotesize #1}}

\def\11{\mbox{$1$}}
\def\pa{\partial}
\def\Cal{\cal}
\def\r{\rho}
\def\s{\sigma}

\def\det{{\rm{det}}}
\def\CH{{\cal{H}}}
\def\CL{{\cal{L}}}
\def\CF{{\cal{F}}}
\def\CK{{\cal{K}}}
\def\ap{{\alpha^\prime}}
\def\DBI{{\rm{DBI}}}
\def\EM{{\rm{EM}}}
\def\D3{{\rm{D3}}}

\def\YM{{\rm{YM}}}

\def\FObig{\left.{}{{\!}^{\!{}^{\!}}_{\!{\!\!}_{}}}\right|_{{}_{\!F\!=0}} }
\def\2{{\geq 2}}

\def\xb{\bar{x}}
\def\yb{\bar{y}}

\def\e{{\mbox{$\varepsilon$}}}
\def\Epsi{{\mbox{\Large $\varepsilon$}}}
\def\E{{\mbox{\Large $\varepsilon$}}}
\def\Eh{{\hat{E}}}
\def\Bh{{\hat{B}}}
\def\thE{{\th^{\varepsilon}}}
\def\La{\Lambda}
\def\SW{\mbox{SW}}
\def\bE{{b^\varepsilon}}
\def\gE{{g^\varepsilon}}

\def\Th{\Theta}
\def\Ga{\Gamma}
\def\GaE{{\Ga^\varepsilon}}
\def\U{\Upsilon}
\def\gEs{{g_s^\varepsilon}}
\def\GEs{{G^{{}^{\mbox{\scriptsize{$\varepsilon$}}}}_s}}
\def\CB{{\Cal{B}}}
\def\CE{{\Cal{E}}}
\def\CD{{\Cal{D}}}
\def\vCB{{\CB}}
\def\vCE{{\CE}}
\def\vCH{{\CH}}
\def\CA{{\Cal{A}}}
\def\ccdot{\!\cdot\!}
\def\Sigmah{{\hat{\Sigma}}}
\def\Hc{{\check{H}}}
\def\SH{{{\Sigma_{\Hc}}}}
\def\BBh{{\hat{\mbox{\boldmath ${B}$}}}}

\def\DDh{{\hat{\mbox{\boldmath ${D}$}}}}
\def\tg{\tan}
\def\GN{Gross-Nekrasov }
\def\q{q}
\def\thth{{{\mbox{\boldmath ${\th}$}}}}
\def\EE{{{\mbox{\boldmath ${\E}$}}}}
\def\1{{-1}}
\def\*{\star}

\newcommand{\eq}{\begin{equation}}
\newcommand{\en}{\end{equation}}
\newcommand{\beeq}{\begin{equation}}
\newcommand{\beqn}{\begin{equation}}
\newcommand{\eeqn}{\end{equation}}
\newcommand{\eneq}{\end{equation}}
\newcommand{\beqr}{\begin{eqnarray}}
\newcommand{\eeqr}{\end{eqnarray}}

\newcommand{\matc}{\begin{array}{c}}
\newcommand{\matcc}{\begin{array}{cc}}
\newcommand{\matccc}{\begin{array}{ccc}}
\newcommand{\matcccc}{\begin{array}{cccc}}
\newcommand{\emat}{\end{array}}

\newcommand{\IH}{\relax{\rm I\kern-.18em H}}
\newcommand{\IR}{\relax{\rm I\kern-.18em R}}
\newcommand{\IK}{\relax{\rm I\kern-.18em K}}
\newcommand{\II}{\hbox{\rm 1\kern-.35em 1}}
\newcommand{\Is}{\relax{\rm 1\kern-.35em 1}}

\begin{document}


\vskip -0.55cm 
\hfill  LMU-TPW-01-04   
\\
\\
\noi{\Large\bf  
Duality Rotations and BPS Monopoles \\with Space and Time Noncommutativity}

\vskip 0.65cm
\noi{\large Paolo Aschieri}\\
{\it Sektion Physik der Ludwig-Maximilians-Universit\"at\\
Theresienstr. 37, D-80333 M\"unchen, Germany\\ 
\small{\it{e-mail address: aschieri@theorie.physik.uni-muenchen.de}}}
\vskip 0.65cm
\noi {ABSTRACT:}
We show that noncommutative electromagnetism and Dirac-Born-Infeld (DBI) 
theory with scalar fields  are $SL(2,R)$ self-dual
when  noncommutativity is light-like and we are in the slowly varying field approximation.
This follows from $SL(2,R)$ self-duality of the commutative DBI Lagrangian and of 
its zero slope limit that we study in detail. 

We study a symmetry of noncommutative static configurations that 
maps space-noncommutativity into space-time (and light-like) noncommutativity.
$SL(2,R)$ duality is thus extended to space-noncommutativity.$\:\,\,$Via 
Seiberg-Witten map we study the nontrivial action of this symmetry 
on commutative DBI theory. In particular space-time noncommutative BPS magnetic monopoles 
corresponds to commutative BPS type magnetic monopoles with both electric and 
magnetic $B$-field background. Energy, charge and tension of these configurations
are computed and found in agreement with that of a D1-string D3-brane system. 
We discuss the dual string-brane configuration.
\vskip 0.20cm
\noi{\small Keywords: Duality; 
Born-Infeld; Noncommutative Gauge Theory; Monopoles }

\vskip 0.50cm

\section{Introduction}
Field theories on noncommutative spaces have received renewed 
interest since their appearance in M-theory and in string theory.  
In particular D-branes effective actions can be described by
noncommutative gauge theories, the noncommutativity arising from 
a nonzero constant background NS $B$ field, 
see \cite{Seiberg:1999vs} and references therein.
When $B\not=0$ the effective physics on the D-brane 
can be described both by a commutative  gauge 
theory $\CL(F+B)$ and by a noncommutative one $\CLh(\Fh)$, where 
\eq
\Fh_{\mu\nu}=\pa_\mu \Ah_\nu - 
\pa_\nu \Ah_\mu -i[ \Ah_\mu\*_{\!\!\!{\textstyle{,}}} \Ah_\nu]
\en 
and  
$\*$ is Moyal star product;  on coordinates
$[x^\mu\*_{\!\!\!{\textstyle{,}}}x^\nu]=
x^\mu\*x^\nu-x^\nu\*x^\nu=i\th^{\mu\nu}$. The noncommutativity parameter 
$\th$ depends on $B$  and on the metric on the D-brane.   
The commutative/noncommutative descriptions are complementary and are 
related by Seiberg-Witten map (SW map). 
Initially  space-noncommutativity has been considered 
($\th^{ij}\not= 0$ i.e. $B_{ij}\not=0$) then  theories also with time 
noncommutativity ($\th^{0i}\not=0$ i.e $B_{0i}\not=0$) have been studied 
\cite{Seiberg:2000gc}.
It turns out that unitarity of noncommutative Yang-Mills theory (NCYM) 
holds only if $\th$ is space-like or light-like 
i.e.$ $ only if the electric and magnetic components of $\th$ (or $B$) are 
perpendicular and if the electric component is not bigger in magnitude
than the magnetic component.
These are precisely the NCYM theories that can be obtained from open 
strings in the decoupling limit 
$\ap\rightarrow 0$ \cite{Aharony:2000gz}. 
In this paper we consider  these two  kinds of space-time 
noncommutativity.
\sk
The effective D-brane actions, for slowly varying fields ($\pa F\sim 0$) 
are respectively commutative and noncommutative Dirac-Born-Infeld 
(DBI)$\,$: 
$\CL_\DBI\stackrel{\sma{\rm{SW map}}}{\longleftarrow\!\!\!\longrightarrow}
\CLh_\DBI$.
In the $\ap\rightarrow 0$ limit \cite{Seiberg:1999vs} 
we obtain respectively a nonlinear commutative $U(1)$ gauge theory and
noncommutative electromagnetism (NCEM); this last is the exact 
effective D-brane action for $\ap\rightarrow 0$.
Some D-brane properties are naturally studied in the noncommutative context
e.g. T-duality \cite{Connes:1998cr, Schwarz:1998qj}, instantons \cite{Nekrasov:1998ss}
and soliton solutions 
(in gauge theory and open string field theory) 
\cite{Gopakumar:2000zd}, \cite{Harvey:2000jt}.
Some other aspects turn out to be easier addressed in the context of 
commutative nonlinear electrodynamics; in particular, in 3+1 dimensions, 
electric-magnetic duality rotations. The general idea is to use SW map in 
order to get new insights about commutative/noncommutative DBI actions
and D-brane physics.
\sk
In the first part of this paper we study duality rotations in  the 
$\ap\rightarrow 0$ limit and away from it, in both cases we see that
noncommutative gauge theory is self-dual if $\th$ is light-like. 
$SL(2,R)$ self-duality of these theories
is shown proving self-duality of the corresponding commutative actions.
In particular we give, to all orders in $\th$, the explicit expression 
of the nonlinear $U(1)$ commutative theory that corresponds 
(for slowly varying fields) to NCEM. 
These results are also generalized to include a scalar field 
interacting with the gauge fields. 
We thus also provide new examples of Lagrangians 
satisfying Gaillard-Zumino self-duality condition.
Formally (covariant derivatives, minimal couplings) NCEM
resembles commutative $U(N)$ YM, and on tori with rational $\th$
the two theories are $T$-dual \cite{Schwarz:1998qj}.
Self-duality of NCEM then hints to a possible duality symmetry 
of the equations of motion of $U(N)$ YM.
Self-duality of NCEM was initially studied in
\cite{Ganor:2000my} to first order in $\th$.
In \cite{Aschieri:2001ks} pure NCEM with zero axion and dilaton 
has been seen to be self-dual under the $SO(2)$ compact subgroup 
of $SL(2,R)$. 
For $\th$ space-like as shown in \cite{Gopakumar:2000na} 
we do not have noncommutative gauge theory self-duality and the S-dual 
of space-like NCYM is a noncommutative open string theory decoupled from 
closed strings. Related work appeared in 
\cite{Rey:2001hh,Lu:2001vv,Alishahiha:2000pu}. 
\sk
Since self-duality holds for $\th$ light-like, it may seem that this 
symmetry has a very restricted range of application.
This is not the case. In the second half of this paper we show that for 
any space-noncommutative static solution we can turn on time-noncommutativity 
and obtain a static solution with light-like noncommutativity. We can then 
apply a duality rotation, switch off the time-noncommutativity and thus 
obtain a new solution of the original pure space-noncommutative theory.
In particular we study BPS solutions of NCDBI and
DBI theories with both electric and magnetic background thus answering the 
following questions:
Consider the commutative BPS solution describing a D1-D3 system in
the presence of a background magnetic field, what happens when we 
turn on a constant electric field? How is this configuration described
in the noncommutative setting?
\sk
There has been quite some activity in the study of space-noncommutative BPS 
states and their relation, via SW map, to commutative BPS states, these studies
have also tested SW map between commutative and space-noncommutative 
gauge theories 
\cite{Seiberg:1999vs,Hashimoto:2000mt,Moriyama:2000mm,Moriyama:2000eh,Hashimoto:2000cf,Hashimoto:2000kq,Hashimoto:2001pc};
in \cite{Mateos:2000qq} pure time-noncommutativity is considered. 
Our work (with space-time noncommmutativity) fits in this context.
In \cite{Gross:2000wc,Gross:2000ph} solutions to the BPS equations of NCEM with just
space-noncommutativity are found, the solution in \cite{Gross:2000wc} describes 
a smeared monopole connected with a string-like flux tube and is interpeted 
as  a D1-string ending on a D3-brane with constant magnetic field background.
In Section 3 we see that this solution remains a BPS solution also when we 
turn on time-noncommutativity. This is due to a symmetry of static 
configurations of NCDBI theory.  Via SW map we obtain the nontrivial action 
of this symmetry in the corresponding commutative DBI theory and show that 
it is a rotation (boost) between the time component $A_0$ of the gauge 
potential and the worldvolume D-brane coordinates $x^i$. 
This boost, first studied in \cite{Gibbons:1998xz}, is similar to the target space 
rotation that relates the linear monopole to the nonlinear monopole 
\cite{Moriyama:2000mm,Hashimoto:2000cf}, \cite{Gibbons:1998xz}. 
We show that this boost maps commutative BPS configurations 
into new BPS configurations; indeed these new configurations correspond 
via SW map to BPS configurations of NCEM, moreover they satisfy a first
order differential equation and their energy equals a
topological charge. In particular we describe explicitly the monopole plus 
string solution in a background that is both electric and magnetic.
The monopole has the fundamental magneton charge and the string tension is
that of a D1-string, this strongly suggests that we have a D1-D3 brane system 
with both electric and magnetic background. The D1-string tension is also 
matched with the string tension of the corresponding space-time 
noncommutative BPS monopole.
We conclude briefly discussing the duality rotated string-brane configuration
that describes an electric monopole in a magnetic background. 
For a different noncommutative approach to describe fundamental strings
see  \cite{Harvey:2000jt}.

\section{Duality rotations in (non)commutative gauge theory}
In this section we first review Gaillard-Zumino 
self-duality condition and self-duality of the (bosonic part of the IIB) 
D3-brane effective Lagrangian 
\cite{Tseytlin:1996it},\cite{Kimura:2001jb}.
We then present two simple arguments showing why duality rotations
require $B$ and $\th$ light-like in the zero slope limit. 
A necessary and sufficient condition for the zero slope limit to exist,
independently form duality rotations, is given and we thus recover the 
condition for the existence of SW map. 
$SL(2,R)$ self-duality of NCDBI and NCEM is then discussed and in the last 
subsection the more general case of DBI theory with an extra scalar field
(Higgs field) is studied. The zero slope of DBI Lagrangian is also 
explicitly given; when no scalar field is present this was given 
for $\th$ light-like in \cite{Aschieri:2001ks}, recently 
it has independently appeared in a different (but for $d=4$ equivalent) 
form in \cite{Garousi:2001jj}.
\sk

\subsection{Self-duality of D3-brane effective action}
\noi Consider in four dimension and with (Einstein)  metric $g_E$ the
Lagrangian density (Lagrangian for short) 
\eq
{\Cal{L}}(F_{\mu\nu}, g_{E\,\mu\nu}, \chi^i)
=\sqrt{g_E}\,
L(F_{\mu\nu}, g_{E\,\mu\nu}, \chi^i)
\en
where $\sqrt{g_E}=\sqrt{-{\rm{det}}g_E}$ and $\chi^i$ are some background fields that can possibly have space-time 
indices. If we define\footnote{The Hodge dual of a two form 
$\Omega_{\mu\nu}$ is defined by 
$\Omega^*_{\mu\nu}\equiv \frac{1}{2}\sqrt{g_E}\,
\epsilon_{\mu\nu\rho\sigma}
\Omega^{\rho\sigma}$, where $\epsilon^{0123}=-\epsilon_{0123}=1$ and
$g_E$ has signature $(-,+,+,+)$. 
We have 
${\Omega^*}^*_{\mu\nu}=-\Omega_{\mu\nu}$,
$\Omega^{*\,\mu\nu}= \frac{1}{2}\sqrt{g_E^{-1}}\, 
\epsilon^{\mu\nu\rho\sigma}\Omega_{\rho\sigma}$.}
 
\beeq
{}~~~~~~~~~~~~~~~~ 
{K^*}^{\mu\nu}~=~\frac{\pa L}{\pa F_{\mu\nu}}~~~~~~~~~~~~~~~~~
\sma{${\left({ {\partial F_{\rho\sigma}}\over{\partial F_{\mu\nu}}  } 
=\delta^{\mu}_{\rho}\delta^{\nu}_{\sigma}-
\delta^{\nu}_{\rho}\delta^{\mu}_{\sigma}\right)}~$}  
\label{defK}
\eneq
then the Bianchi identity and the equations of motion (EOM) for $\CL$ 
read
\beqr
\begin{array}{c} 
\pa_{\mu}(\sqrt{g_E}\, F^{*\,\mu\nu} )=
{\pa}_{\mu}
\tilde{F}^{\mu\nu}  =0~,\\[0.8em]
{\pa}_{\mu}(\sqrt{g_E}\,  K^{*\,\mu\nu} )=
{\pa}_{\mu}
\tilde{K}^{\mu\nu}  =0~,
\end{array}\label{3}
\eeqr
where ${\tilde{F}}^{\mu\nu}
\equiv 
\frac{1}{2}\epsilon^{\mu\nu\rho\sigma} F_{\rho\sigma}$ and 
${\tilde{K}}^{\mu\nu}
\equiv 
\frac{1}{2}\epsilon^{\mu\nu\rho\sigma} K_{\rho\sigma}$.
Under the
infinitesimal transformations 
\beeq
\delta
\left(
\matc
K \\
F
\emat
\right)
=
\left(
\matcc
\mbox{A} & \mbox{B} \\
\mbox{C} & \mbox{D}
\emat
\right)
\left(
\matc
K \\
F
\emat
\right)~\label{FKtrans}
\eeqn
\beeq
\!\!\!\!\!\!\!\!\!\!\!\!\!
\!\!\!\!\!\!\!\!\!\!\!\!\!\!\!\!\!\!\!\!\!\!\!\!\!\!\!\!\!\!\!\!\!\!\!\!\!
\delta \chi^i=\xi^i(\chi)~,
\label{transchi}
\eeqn
the $\CL^{{}^{{}_{\scriptstyle g_{{\!}_E} ,\chi}}}$ EOM (\ref{3}) are 
mapped into the
$\CL^{{}^{{}_{\scriptstyle g_{{\!}_E},\chi+\del\chi}}}$ EOM. 
Consistency of (\ref{FKtrans}),(\ref{transchi}) 
with the definition of $K$, i.e.
$\tilde{K}+\del\tilde{K}=\frac{\pa}{\pa(F+\del F)}\CL(F+\del F,g_E,\chi+\del\chi)$ 
holds in particular if 
$~\!
\left( {}^{\mbox{\sma {A}} 
\mbox{\sma {B}}}_{\mbox{\sma {C}} 
\mbox{\sma{D}}} \right)$ 
belongs to the Lie algebra of $SL(2,R)$,
and the variation of the Lagrangian under 
(\ref{FKtrans}),(\ref{transchi})
can be written as
\eq
\delta {\Cal{L}}\equiv(\delta{_F}+\delta{_\chi}){\Cal{L}}
={1\over 4}(\mbox{B}F\tilde{F}+\mbox{C}K\tilde{K})
 \label{cond}~,
\en
where $\del_F\CL=\frac{1}{2}\del F \frac{\pa \CL}{\pa F}\,,~
 \del_\chi\CL=\del \chi^i \frac{\pa \CL}{\pa \chi^i}$ \cite{Gaillard:1981rj}.
When (\ref{cond}) holds,  under a finite $SL(2,R)$ rotation
a solution $F$ of $\CL^{{}^{{}_{\scriptstyle g_{{\!}_E},\chi}}}$ is mapped
into a solution $F'$ of $\CL^{{}^{{}_{\scriptstyle g_{{\!}_E},\chi'}}}$ 
and we say that $\CL^{{}^{{}_{\scriptstyle g_{{\!}_E},\chi}}}$
is self-dual; notice that the (Einstein) metric $g_E$ is invariant. 
A self-duality condition equivalent to (\ref{cond}) is obtained using 
(\ref{FKtrans}) to evaluate the $\del_F\CL$ term in (\ref{cond})
\eq
\del_\chi\CL=\frac{1}{4}{\rm{B}}F\tilde{F}
           -\frac{1}{4}{\rm{C}}K\tilde{K}
           -\frac{1}{2}{\rm{D}}{F}\tilde{K}
\label{cond1}~~.
\en
If also the $\chi$ background fields are invariant under duality rotations 
then the maximal duality group is $U(1)$ 
\cite{Gaillard:1981rj}, see also the nice review \cite{Kuzenko:2001uh}.
Viceversa we can always extend a $U(1)$ self-dual Lagrangian to a 
$SL(2,R)$ self-dual one introducing two real valued scalars 
$S=S_1+iS_2$ (axion and dilaton) 
\cite{Gibbons:1996ap,Gaillard:1997zr,Gaillard:1997rt}

Among the properties of self-dual theories we recall that 
\noi
{\it{i}}) self-duality 
implies invariance under 
Legendre transformations with respect to the field 
strength \cite{Gaillard:1997rt,Kuzenko:2001uh};\\
{\it{ii}}) Consider a parameter or field $\lambda$ invariant under 
the transformations (\ref{FKtrans}), (\ref{transchi}) 
then, if $\xi^i$ is independent of $\lambda$, the derivative 
$\frac{\pa}{\pa{\lambda}}{\Cal{L}}(F,g_E,\chi,\lambda)$
is also invariant under  
(\ref{FKtrans}),(\ref{transchi})\footnote{Proof \cite{Gaillard:1981rj}: 
differentiate w.r.t. $\pa_\la\equiv\frac{\pa}{\pa\la}$
the expression $\del\CL=\del_\chi\CL+\del_F\CL=
\del_\chi\CL+\frac{1}{2}\del F \frac{\pa\CL}{\pa F}$ 
to obtain 
$\pa_\la\del\CL=\del_\chi\pa_\la\CL+\frac{1}{2}\del F\pa_\la
\frac{\pa\CL}{\pa F}+\frac{1}{2}\pa_\la(\del F)
\frac{\pa\CL}{\pa F}=\del\pa_\la\CL+\frac{1}{2}\mbox{C}(\pa_\la K)\tilde{K}$
i.e. $\del\pa_\la\CL=\pa_\la(\del\CL-\frac{1}{4}\mbox{C}K\tilde{K})=
\pa_\la
(\del\CL-\frac{1}{4}\mbox{C}K\tilde{K}-\frac{1}{4}\mbox{B}F\tilde{F})=0$
where we used (\ref{cond}) in the last equality.}.
In particular since the metric is invariant then 
the energy momentum tensor is invariant.

Property {\it{ii}}) provides a way to construct new self-dual Lagrangians 
from known ones. Indeed given the self-dual lagrangian density 
${\Cal{L}}(F,\mbox{g}_E,\chi)$
consider the new lagrangian density 
\beqn
\check{{\Cal{{L}}}}(F,{g}_E +f(\phi),\chi)\equiv 
{\Cal{L}}(F,\mbox{g}_E,\chi)\label{L+higgs}
\eeqn
where ${g}_{E\,\mu\nu}+f_{\mu\nu}(\phi_E)=\mbox{g}_{E\mu\nu} $ and 
$f_{\mu\nu}(\phi_E)$ is a function of the metric  ${g}_E$, 
the scalar fields $\phi_E^{a}$ and their derivatives 
{e.g.}  $f_{\mu\nu}(\phi_E)=
\sum_{a}\pa_{\mu}\phi_E^{a}\pa_{\nu}\phi_E^{a}$.
Then $\check{\Cal{L}}$ is self-dual under  (\ref{FKtrans}), 
(\ref{transchi})
and $~\delta {g}_E=\delta\phi_E^{a}=0~$ indeed
$$
(\del_F+\del_{\chi}+\del_{\phi_E}) \check{\Cal{L}}=
(\del_F+\del_{\chi})\check{\Cal{L}}=(\del_F+\del_{\chi}){\Cal{L}}=
{1\over 4}(\mbox{B}F\tilde{F}+\mbox{C}K\tilde{K})=
{1\over 4}(\mbox{B}F\tilde{F}+\mbox{C}\check{K}\tilde{\check{K}})~
$$
in the last line we used 
$\tilde{\check{K}}{}^{\mu\nu}\equiv\pa{\check{\Cal{L}}}/
\pa{F_{\mu\nu}}=\pa{\Cal{L}}/\pa{F_{\mu\nu}}=\tilde{K}^{\mu\nu}$.
Concerning the $\phi_E^a$ EOM
$$
\frac{\pa}{\pa \phi_E^a}\check{\Cal{L}} -
\pa_{\mu}\frac{\pa}{\pa\pa_{\mu}\phi_E^a}\check{\Cal{L}}=0 $$
%
%
they do not change under the duality rotation because [recall {\it{ii}})]
$\phi_E^a$ 
is invariant and $\xi^i(\chi)$ is $\phi_E^a$ independent.
\sk

We now discuss self-duality of 
the D3-brane effective action in a IIB supergravity background 
with constant NS-NS two-form. 
The background two-form can be gauged away in the bulk and we are left
with the field strength $\CF=F+B$ on the D3-brane. 
Here $B$ is defined as the constant part of $\CF$, 
or $B\equiv\CF|_{\rm{spatial}\, \infty}$ since 
$F$ vanish at spatial infinity.
{}For slowly varying fields
the Lagrangian is a generalized version of Dirac-Born-Infeld Lagrangian,
in string and in Einstein frames, it respectively reads:
\beqr
\CL_{\rm{DBI}}&=&
\frac{-1}{\ap^2 g_s}\sqrt{-{\rm{det}}(g+\ap {\CF})}
+\frac{1}{4}C{\CF}\tilde{\CF}\nonumber\\[1.3em]
\!\!\!&=&\!\!\frac{-1}{\ap^2}\sqrt{-{\rm{det}}(g_E+\ap S_2^{1/2}{\CF})}
+\frac{1}{4}S_1{\CF}\tilde{\CF}\label{LDBIWZ}\\[1.3em]
\!\!\!&=&\!\!\frac{-1}{\ap^2}\sqrt{-g_E}
\; {\sqrt{1+\frac{\ap^2}{2} S_2 \CF^{\,2}-\frac{\ap^4}{16}S_2^2(\CF\CF^*)^2}}+
\frac{1}{4}S_1{\CF}\tilde{\CF}\nonumber
\eeqr
where in the second line
$g_E=g_s^{-\frac{1}{2}}g$ and $S=S_1+iS_2=C+\frac{i}{g_s}\;$
while,
in the last line, we have simply expanded the 4x4 determinant and 
$\CF^{\,2}\equiv\CF_{\mu\nu}\CF^{\mu\nu}\,,~
\CF\CF^*\equiv\CF_{\mu\nu}\CF^{*\,\mu\nu}$.
Also, with abuse of notation, we have set $2\pi \ap=\ap,~2\pi g_s=g_s$.

Under the $SL(2,R)$ rotation
\beeq
\left(
\matc
\CK^\prime \\
\CF^\prime
\emat
\right)
=
\left(
\matcc
a & b \\
c & d
\emat
\right)
\left(
\matc
\CK \\
\CF
\emat
\right)~~,~~~~
\left(
\matcc
a & b \\
c & d
\emat
\right)
\in SL(2,R)~,\label{F+Brot}
\eeqn
\beqn
{}~~~S^\prime=\frac{aS+b}{cS+d}\label{Srot}~,~~g_E^\prime=g_E~,~~(\ap)'=\ap
\eeqn
where  $\tilde{\CK}=\frac{\pa}{\pa \CF}\CL$, the Lagrangian  $\CL_{\rm{DBI}}$
satisfies the self-duality condition
(\ref{cond1}) with $F$ and $K$ replaced by $\CF$ and $\CK$. [The infinitesimal
transformations (\ref{FKtrans}),(\ref{transchi}) can be obtained from the
finite transformations (\ref{F+Brot}),(\ref{Srot}) using 
$\left({}^a_c~{}^b_d\right)\approx 
\left({}^{{}_{{}^{{}^{{}_{{}_{{}_{{}_{\scriptstyle 
1}}}}}}}}_{{}^{{}_{{}_{\scriptstyle 0}}}}~\!
{}^{{}_{{}^{{}^{{}_{{}_{{}_{{}_{\scriptstyle 
0}}}}}}}}_{{}^{{}_{{}_{\scriptstyle 1}}}}\!\right)+
\left({}^{A}_C~{}^B_{D}\right)$].
{}For completeness we should add to $\CL_\DBI$ the Wess-Zumino term
$\tilde{C}_4=\frac{1}{24}
\epsilon^{\mu\nu\r\s}C_{\mu\nu\r\s}$, where $C_{\mu\nu\r\s}$
is the pull-back on the D3-brane of the RR four-form $C_4$. Under the 
$SL(2,R)$ action $C_4$ is invariant so that 
$\CL_\D3=\CL_\DBI+\tilde{C}_4$ is self-dual. Moreover under $SL(2,R)$ 
the $g_E, S, C_4$ ten dimensional IIB supergravity 
background (with zero NS-NS and RR two-forms) is rotated into the 
$g_E, S', C_4$ one (with zero NS-NS and RR two-forms)
\cite{Schwarz:1983wa},\cite{Kimura:2001jb}.

The  explicit expression of $\CK$ is 
\eq
{\CK}_{\mu\nu}=
\frac{S_2\CF^*_{\,\mu\nu}+\frac{\ap^2}{4}S_2^2\, \CF\CF^*\,\CF_{\mu\nu}}{
\sqrt{1+\frac{\ap^2}{2} S_2 \CF^2-\frac{\ap^4}{16}S_2^2(\CF\CF^*)^2}}
+S_1\CF_{\mu\nu}~.
\en
{}Consider now a constant dilaton and axion $S$, then
$S_2^{-1}=g_s=g^2_\YM$ 
plays the role of gauge coupling constant and 
the term $S_1\CF\tilde\CF$  becomes a total derivative that does 
not affect the $\CF$ EOM (but is needed in $\CK$).
From (\ref{F+Brot}), one can extract how $B$ 
(the value of $\CF$ at spatial infinity) transforms
\eq
{B'}_{\mu\nu}=(cS_1+d)B_{\mu\nu} + c
\frac{S_2 B^*_{\,\mu\nu}+\frac{\ap^2}{4} S_2^2\, B B^*\, B_{\mu\nu}}{
\sqrt{1+\frac{\ap^2}{2} S_2 B^2-\frac{\ap^4}{16}S_2^2(B B^*)^2}}~;\label{Brot}
\en
this transformation is independent from the slowly varying fields 
approximation.
\sk 
We have seen self-duality of $\CL_\DBI$ and of $\CL_\D3$ under
$\CF, \CK$ rotations. Let's now consider $F$, instead of  $\CF$, as the 
dynamical field in $\CL_\DBI(F+B,g_E,S)$. This Lagrangian 
contains constant terms and terms linear in $F$ that are total derivatives, 
we drop them and obtain the Lagrangian
\eq
\CL_\2\equiv\CL_\DBI-\CL_0-\CL_1\equiv
\CL_\DBI-\CL_\DBI\FObig -\frac{1}{2}\frac{\pa\CL_\DBI}{\pa F}\FObig F
\label{DBIr}
\en
that is at least quadratic in $F$. We also split $\tilde\CK=\frac{\pa}{\pa \CF}\CL=\frac{\pa}{\pa F}\CL$
as 
\eq
\tilde\CK={\frac{\pa\CL_1}{\pa F}}+\tilde{K}~~,~~~{\rm{where}}~~ 
\tilde{K}\equiv\frac{\pa \CL_\2}{\pa F}~.
\en
In order to show self-duality of
$\CL_\2(F,g_E,B,S)$ w.r.t. $F$ we extract from (\ref{cond1}) 
[i.e. $\del_S\CL=\frac{1}{4}{\rm{B}}\CF\tilde\CF
           -\frac{1}{4}{\rm{C}}\CK\tilde\CK
           -\frac{1}{2}{\rm{D}}\CF\tilde\CK$]
the terms that contains powers of $F$ of order $\geq 2\, ,$ and obtain
\eq
\del_S\CL_\2=\frac{1}{4}{\rm{B}}F\tilde{F}
-\frac{1}{4}{\rm{C}}{K}\tilde{K}
-\frac{1}{2}{\rm{D}}F\tilde{K}
-\frac{1}{2}(-{\rm{C}}{\widetilde{\frac{\pa\CL_1}{\pa F}}}+{\rm{D}}B) 
(\tilde{K}-\frac{1}{2}\frac{\pa \tilde{K}}{\pa F}\FObig F)\label{5}~.
\en
We now disentangle in 
$\frac{\pa}{\pa F}\CL_\DBI=
\frac{\pa}{\pa B}\CL_\DBI$ 
the terms of order $0$ in $F$ from all the other terms and obtain
$$\frac{\pa \CL_1}{\pa F}=\frac{\pa \CL_0}{\pa B}~~~~~\mbox{ and } 
~~~~~\tilde{K}=\frac{\pa \CL_1}{\pa B}+ \frac{\pa \CL_\2}{\pa B} =
\frac{1}{2}\frac{\pa \tilde{K}}{\pa F}\FObig F
+ \frac{\pa \CL_\2}{\pa B} ~~. $$
Relation (\ref{5}) then becomes the self-duality equation for $\CL_\2$
\eq
(\del_B+\del_S)\CL_\2=\frac{1}{4}{\rm{B}} F\tilde{F}
           -\frac{1}{4}{\rm{C}}{K}\tilde{K}
           -\frac{1}{2}{\rm{D}}{F}\tilde{K}~.\label{6}
\en


\subsection{Open/closed strings and light-like noncommutativity}

\noi The open and closed string parameters are related by 
(see \cite{Seiberg:1999vs}, 
the expressions for $G$ and $\th$ first appeared in 
\cite{Chu:1999qz})
\eq
\begin{array}{l}
\displaystyle
\frac{1}{g+\ap B}=G^{-1}+\frac{\th}{\ap} ~~\nonumber\\[1em]
\displaystyle
g^{-1}=(G^{-1}-\th/\ap)\,G\,(G^{-1}+\th/\ap)=G^{-1}-\ap^{-2}
\th\, G\, \th\nonumber\\[1em]
\displaystyle
\ap B=-(G^{-1}-\th/\ap)\,\th/\ap\,(G^{-1}+\th/\ap)\nonumber\\[1em]
\displaystyle
G_s=g_s 
\sqrt{\frac{{\rm{det}}G}{\det(g+\ap B)}}=g_s\sqrt{\det G\;
\det{(G^{-1}+\th/\ap)}}=g_s\sqrt{\det g^{-1}\;
\det{(g+\ap B)}}
\end{array}
\label{Gsgs}
\en 
If we define $\Omega\equiv-G\frac{\th}{\ap^2} G$ then $g,B$ 
are related to $G,\Omega$ via the change of coordinates
\eq
\!\!\begin{array}{c}
G=M^T g M\\
\Omega=M^T B M
\end{array}
{}~~~~\mbox{ with }~~  M=g^{-1}(g+\ap B)=G^{-1}(G+\ap\Omega)=(G^{-1}-\th/\ap)G~.
{}~\label{M}
\en
The decoupling limit 
$\ap\rightarrow 0$ with $G_s,G,\th$ 
nonzero and finite 
\cite{Seiberg:1999vs}  
leads to a well defined field theory only if
$B$ is space-like or
light-like \cite{Aharony:2000gz}. 
Looking at the closed and open string coupling constants it is 
easy to see why one needs this space-like or light-like condition on 
$B$.
Consider the coupling constants ratio $G_s/g_s$, that
expanding the 4x4 determinant reads 
(here $B^2=B_{\mu\nu}B_{\r\s}g^{\mu\r}g^{\nu\s}$, 
$\th^2=\th^{\mu\nu}\th^{\r\s}G_{\mu\r}G_{\nu\s}$ and so on)
\eq
\frac{G_s}{g_s}=\sqrt{1+\frac{\ap^{-2}}{2} 
\th^2-\frac{\ap^{-4}}{16}(\th \th^*)^2} 
\;=\;
\sqrt{1+\frac{\ap^2}{2} B^2-\frac{\ap^4}{16}(B B^*)^2}\label{Gs/gs}~.
\en
Both $G_s$ and $g_s$ must be positive; since $G$ and $\th$ are by
definition 
finite for $\ap\rightarrow 0$ this implies 
$\th \th^*=0$ and $\th^2\geq 0$. Now  $\th \th^*=0 \Leftrightarrow 
\det\th=0 \Leftrightarrow \det B=0 \Leftrightarrow B B^*=0$.
In this case from (\ref{Gs/gs}) we also have $\th^2=\ap^4B^2$.
In  conclusion a necessary condition for the $\ap\rightarrow 0$ 
limit to be well defined is
\eq 
B^2\geq 0~,~~B B^*= 0
~~~~~~~\mbox{i.e.}~~~~~~~\th^2\geq 0~,~~\th \th^*=0~~.
\label{Bth}
\en
This is the condition for $B$ (and 
$\th$) to be space-like or light-like.
Indeed with Minkowski metric (\ref{Bth}) reads 
$\vec{B}^2-\vec{E}^2\geq 0$ and $\vec{E}\perp\vec{B}$. 
In the next subsection we see that a necessary and sufficient 
condition for the $\ap\rightarrow 0$ limit to be well defined
is (\ref{Bth}) and 
\eq
1-\frac{1}{2}\th F\not=0 \label{acond}~~.
\en 
Notice that space time is divided in the regions
$1-\frac{1}{2}\th F>0$ and $1-\frac{1}{2}\th F<0$. We generically
have $1-\frac{1}{2}\th F>0$ because we have considered $F=0$ at 
spatial infinity. Condition (\ref{acond}) is the condition 
for SW map to be well defined \cite{Seiberg:1999vs}, 
see also \cite{Jurco:2001my}, 
indeed for constant $F$ it is equivalent to the existence 
of $\Fh_{\mu\nu}=[\frac{1}{1+F \th }F]_{\mu\nu}$.

\sk

We now require the $\ap\rightarrow 0$ limit to be compatible with 
duality rotations. From (\ref{Bth}) we immediately see that we have 
to consider only the light-like case $B^2=  B B^*= 0$;
indeed under $SL(2,R)$ rotations 
the electric and magnetic fields mix up;
in particular under an $SL(2,R)$  rotation 
$\left({}^a_c{~}^b_d\right)$ with $d=-cS_1$,
a space-like $B$ becomes time-like.

A complementary perspective that leads to the light-like condition
$B^2=B B^*=0$ is as follows.
We fix $G_s,G,\th$ and rescale $\ap\rightarrow 
\epsilon\ap$. This defines the $\ep$ dependence of the open string 
variables: $g_s(\ep),g(\ep),B(\ep)$. 
Under a duality rotation we obtain $g'_s(\ep),g'(\ep),B'(\ep)$.
The two operations of duality rotation and $\ap$ rescaling are 
compatible if they commute, i.e. the open
string variables $G'_s,G',\th'$ obtained from $g'_s(\ep),g'(\ep),B'(\ep)$
must be $\ep$ independent. This is the case if $\th$ (or $B$) 
is light-like.
As shown in \cite{Gopakumar:2000na} if $B$ is space-like and $S_1=0$, 
under a $\pi/2$ rotation $G'$ depends linearly 
on $\ep^{-1}$; letting $\ep\rightarrow 0$ open strings decouple 
from closed strings but massive open string states do not decouple 
from massless open string states so that the S-dual of this NCEM is 
a noncommutative open string theory.

\subsection{Self-duality of NCDBI and NCEM}

\noi We now study duality rotations for noncommutative 
Dirac-Born-Infeld (NCDBI) theory and its zero slope limit that is
noncommutative electromagnetism (NCEM).
The relation between NCDBI and DBI Lagrangians is \cite{Seiberg:1999vs}
\eq
\CLh_\DBI(\Fh,G,\th,G_s,C)=\CL_\DBI(F+B,g,g_s,C)+O(\pa F)+\rm{tot.der.}
\label{NCDBI}
\en
where $\Fh$ is the noncommutative $U(1)$ field strength
and $O(\pa F)$ stands for higher order derivative corrections
(these corrections are at least quadratic in $\pa$).
The NCDBI Lagrangian is: 
\eq
\CLh_\DBI(\Fh,G,\th,G_s,C)=
\frac{-1}{\ap^2 G_s}\sqrt{-{\rm{det}}(G+\ap {\Fh})}
+\frac{1}{4}C\Fh\widetilde{\Fh} +O(\pa\Fh)~~;
\en
as in the commutative case also $C\Fh\widetilde{\Fh}$ is a total derivative.

In the slowly varying field approximation the action of duality 
rotations on $\CLh_\DBI$  
is derived from self-duality of $\CL_\DBI$. 
If $\Fh$ is a solution of the 
$\CLh_\DBI^{G,\th,G_s}$ EOM then $\Fh'$ obtained via 
$\Fh\stackrel{\sma{\rm{SW map}}}{\longleftarrow\!\!\!\longrightarrow}
\CF\stackrel{\sma{\rm{duality\, rot.}}}{\longleftarrow\!\!\longrightarrow}
\CF'
\stackrel{\sma{\rm{SW map}}}{\longleftarrow\!\!\!\longrightarrow}
\Fh'$ is a solution of the 
$\CLh_\DBI^{G',\th',G^\prime_{\!s}}$ EOM where $G',\th',G^\prime_s$
are obtained using (\ref{Gsgs}) from  $g'=S'^{-\frac{1}{2}}_2 g_E$
and $B'$. 
More precisely in (\ref{NCDBI}) we can replace
$\CL_\DBI(F+B,g_E,S)$  with (\ref{DBIr}) that as shown in (\ref{6})
is self-dual. Using
\eq
\frac{1}{g_s}\sqrt{-{\rm{det}}(g+\ap\CF)}
=\frac{\sqrt{G}}{G_s}\sqrt{\frac{{\rm{det}}(g+\ap B+\ap{F})}{{
\rm{det}}(g+\ap B)}}=
\frac{1}{G_s}\sqrt{-{\rm{det}}(G+\ap{F}+ G\th F)}
\label{LGth}
\en
we then rewite $\CL_\2(F,g_E,B,S)$ 
in terms of the open string parameters $G,\th,G_s,C$
\eq
\CL_\2(F,G,\th,G_s,C)=
\frac{-1}{\ap^2 G_s}\sqrt{-{\rm{det}}(G+\ap{F}+ G\th F)}+
\frac{\sqrt{G}}{\ap^2 G_s}(1-\frac{1}{2}\,\th F)\label{Lgeq2}
\en
This lagrangian satisfies the self-duality condition (\ref{cond}) with 
$\del\CL_\2=(\del_F+\del_{G}+\del_\th+\del_{G_s}+\del_C)\CL_\2$.
Via 
$\Fh\stackrel{\sma{\rm{SW map}}}{\longleftarrow\!\!\!\longrightarrow}
F\stackrel{\sma{\rm{duality\, rot.}}}{\longleftarrow\!\!\longrightarrow}
F'\stackrel{\sma{\rm{SW map}}}{\longleftarrow\!\!\!\longrightarrow}
\Fh'$
we then map solutions $\Fh$ of the $\CLh_\DBI^{G,\th,G_s}$ EOM 
to solutions $\Fh'$ of the $\CLh_\DBI^{G',\th',G^\prime_{\!s}}$ EOM.
(The non uniqueness of SW map is not an issue here because SW map
is unique up to gauge transformations).
\sk
In the light-like case relations (\ref{Gsgs}) simplify considerably.
The open and closed string coupling constants coincide 
so that $S=S_1+iS_2=C+\frac{i}{g_s}=C+\frac{i}{G_s}\,$.
Use of the relations
\eq
\Omega^*_{\mu\r}{\Omega^*}^{\r\nu}-
\Omega_{\mu\r}\Omega^{\r\nu}=
\frac{1}{2}\Omega^2\,\del_{\mu}^{~\nu}~\,,\,~~ 
\Omega_{\mu\r}{\Omega^*}^{\r\nu}=
{\Omega^*}_{\mu\r}\Omega^{\r\nu}=
\frac{-1}{4}\Omega\Omega^*\,\del_\mu^{~\nu}\label{S}
\en
valid for any antisymmetric tensor $\Omega$, shows that 
any two-tensor at least cubic in $\th$ (or $B$) vanishes.
It follows that $g^{-1}G\,\th=\th$ and that the 
raising or lowering of the  
$\th$ and $B$ indices is independent from the metric used.
We also have
\eq
{}~~~~~~~~~~~B_{\mu\nu}=-{\ap}^{-2}\th_{\mu\nu}\label{Btheta}~~~~ 
{\rm{\sma{~~~~~~(in~ string ~frame)}}}
\en
In the light-like case 
the  $B$ rotation (\ref{Brot}) simplifies to
\eq
B^\prime_{\mu\nu}=(cS_1+d)B_{\mu\nu}+cS_2B^*_{\,\mu\nu}~.\label{Bsrot}
\en
[we do not specify whether ${B^*}_{\mu\nu}$ (or ${\th^*}^{\mu\nu}$) 
is obtained with 
Einstein or string metric because  ${B^*}_{\mu\nu}$ 
(or ${\th^*}^{\mu\nu}$) is the same
in both metrics].
In order to describe the $SL(2,R)$ action on the open string variables
we express them in Einstein frame
and then use (\ref{Srot}) and (\ref{Bsrot}) to obtain
\eq
G_E^\prime=G_E~~,~~~\th^{\prime\,\mu\nu}=
\frac{(cS_1+d)\th^{\mu\nu}+cS_2{\th^*}^{\mu\nu}}{|cS+d|^2}~~~~~~~~
{\rm{i.e.}}~~~
\th^\prime+i\th^{\prime *}
=\frac{\th+i\th^{*}}{cS+d}\label{GETrot}
\en
Therefore, for $\th$ light-like, $G_E$ is invariant under duality rotations,
(\ref{Lgeq2}) is self-dual and NCDBI Lagrangian is self-dual. In particular 
if $S_1=0$, under a $\pm\pi/2$ rotation $S_2\rightarrow -{\frac{1}{S_2}}$
and $\th\rightarrow \pm{\frac{1}{S_2}}\th^*$.
\sk

We now consider the decoupling limit. 
The determinant in (\ref{LGth}) can be evaluated as sum of
products of traces (Newton-Leverrier formula). Each trace can 
then be rewritten in terms of the six basic Lorentz invariants
$F^2,~F F^*,~F\th,~F\th^*,~\th^2$, and $\th\th^*=0$ because of (\ref{Bth}). 
Explicitly 
$$
\begin{array}{l}
{\!\!\!\!\!\!}{\rm{det}}{G^{-1}}\,{\rm{det}}(G+\ap{F}+G\th F)=\nonumber\\
{}~~~~~~~~~~~~~
(1-\frac{1}{2}\th F)^2+\ap^2[\frac{1}{2}F^2+\frac{1}{4}
\th F^*\;FF^*-\frac{1}{32}\th^2(FF^*)^2]-\ap^4(\frac{1}{4}FF^*)^2
\end{array}
$$
Finally we take the $\ap\rightarrow 0$ limit of 
(\ref{Lgeq2}). This limit is well defined iff
$\,1-\frac{1}{2}\th F>0$, if  $\,1-\frac{1}{2}\th F<0$ we obtain also an
infinite total derivative addend that we can neglect; with this prescription 
the $\ap\rightarrow 0$  limit is well defined iff 
$\,1-\frac{1}{2}\th F\not=0$. 
We denote by $\CL_{\ap\!\!\rightarrow 0}$ the resulting 
Lagrangian, in string and Einstein frame we respectively have
%
\beqr
\CL_{\ap\!\!\rightarrow 0}&=&
\frac{\sqrt{G}}{G_s}\;
\frac{-\frac{1}{4}F^2-
\frac{1}{8}\th F^*\,FF^*+\frac{1}{64}\th^2 (F F^*)^2}{|1-
\frac{1}{2}\th F|}\, +\frac{1}{4}C F\tilde{F}\label{L}\\[1.1em]
&=&\sqrt{G_{\!E}}\;S_2\;\frac{-\frac{1}{4}F^2-
\frac{1}{8}\th F^*\,FF^*+\frac{1}{64}\th^2 (F F^*)^2}{|1-
\frac{1}{2}\th F|}\, +\frac{1}{4}S_1F\tilde{F}~\,.\nonumber
\eeqr
We thus have an expression for NCEM in terms of 
$G,\th,G_s,C$ and $F$  
\beqr
\widehat{\CL}_{\rm{EM}}
&\equiv&\frac{\sqrt{G}}{G_s}\;\frac{-1\,}{4}{\Fh}^2\, +\,\frac{1}{4}C 
F\tilde{F}\\[1.1em]
&=&\CL_{\ap\!\!\rightarrow 0}+O(\pa F)+\rm{tot. ~der.}\nonumber
\eeqr
\sk
In the light-like case we have $\th^2=0$ and 
one can directly verify that the Lagrangian (\ref{L}) 
when
$1-\frac{1}{2}\,\th F>0$
satisfies the self-duality condition (\ref{cond1}) with $\chi=\th,S$ 
so that (in the slowly varying field approximation) NCEM is self-dual.
More easily, self-duality of (\ref{L}) when $1-\frac{1}{2}\,\th F>0$
follows from self-duality of $\CL_\2(F,G,\th,G_s,C)$; indeed, 
for $\th$ light-like we have seen that the $\ap\rightarrow 0$ limit 
is well defined and compatible with duality rotations
\footnote{Self-duality in the region $1-\frac{1}{2}\,\th F<0$
holds too. Because of the absolute value the first addend in the 
Lagrangian gains a minus sign. This new Lagrangian satisfies the 
self-duality condition (\ref{cond1}) under the $SL(2,R)$ rotation
(\ref{FKtrans}),(\ref{Srot}), $G'_E=G_E$ and 
$\th^\prime+i\th^{\prime *}
=\frac{\th+i\th^{*}}{c\,{\overline{S}}^{{}^{{}^{}}}+d}$.
Proof: If $\CL(F,\th,S)$ satisfies (\ref{cond1})
under (\ref{FKtrans}) and (\ref{transchi}) that we write more explicitly as
$\delta S=\delta_{\left({}^A_{C}{}^{B}_{D}\right)}S$ and
$\delta\th =\delta_{\left({}^A_{C}{}^{B}_{D}\right);S}\th$, 
then $-\CL(F,\th,S)$ satisfies (\ref{cond1}) under (\ref{FKtrans})
and 
$\delta_{\left({}^{~A}_{-C}{\!}^{\!-B}_{~D}\right)}S$ and
$\delta_{\left({}^{~A}_{-C}{\!}^{\!-B}_{~D}\right);S}\th$.
Similarly for $-\CL(F,\th,-\bar{S})$. Finally notice that
$\delta_{\left({}^{~A}_{-C}{\!}^{\!-B}_{~D}\right)}(-\bar{S})$ 
is equivalent to $\delta_{\left({}^A_{C}{}^{B}_{D}\right)}S$.
About the new $\th$ transformation rule we recall that the original
$\th$ transformation was found setting $F=0$ i.e. was found in the
region $1-\frac{1}{2}\,\th F>0$, there is no reason why it should hold
also in the $1-\frac{1}{2}\,\th F<0$ region.}.
\sk
\subsection{Dirac-Born-Infeld theory with extra scalar fields}

\noi 
The four dimensional DBI Lagrangian with scalar fields (Higgs fields) 
$\phi^a$, $a=1,...n$, can be obtained from the $4+n$ dimensional pure DBI 
Lagrangian where the gauge fields $A_M=(A_\mu,A_a)$ depend only on the 
$x^\mu$ coordinates $\mu=0,...3$, the metric $g_{MN}$ satisfies $g_{\mu b}=0$,
$g_{ab}=\delta_{ab}$, the $B_{MN}$ field satisfies $B_{\mu b}=B_{ab}=0$
and $\phi_a=\ap A_a$ \cite{Gibbons:1998xz}.

This derivation applies also in the noncommutative case, here $G_{MN}$ 
satisfies $G_{\mu b}=0$,
$G_{ab}=\delta_{ab}$. Relation (\ref{Gsgs}) then implies $\th^{\mu b}=
\th^{ab}=0$. We also set $\phih_a=\ap \Ah_a$,  
$\Dh_\mu\phih_a=\pa_\mu\phih_a-i[\Ah_\mu\star_{\!\!\!\displaystyle ,}
\phih_a]$ and consider $[\phih_a\star_{\!\!\!\displaystyle ,}
\phih_b]=0$. The fields $\Ah_\mu,\phih_a$ are related to the $A_\mu, \phi^a$
fields via the $4+n$ dimensional SW map.
We have
\beqr
\!\!\!\!\!\!\!\!\!\CL_{\rm{DBI}}(\CF,\phi,g,g_s,C)&=&
\frac{-1}{\ap^2 g_s}\sqrt{-{\rm{det}}(g+\ap {\CF}+\ap^2\pa\phi_a\pa\phi^a)}
+\frac{1}{4}C{\CF}\tilde{\CF}\label{DBIH}\\[1em]
\!\!\!\!\!\!\!\!\!\CLh_\DBI(\Fh,\phi,G,\th,G_s,C)&=&
\CL_\DBI(F+B,\phi,g,g_s,C)+O(\pa F, \pa\pa \phi)+\rm{tot.der.}
\label{Comm;NCDBIH}\\[1em]
\!\!\!\!\!\!\!\!\!\CLh_\DBI(\Fh,\phih,G,\th,G_s,C)&=&
\frac{-1}{\ap^2 G_s}\sqrt{-{\rm{det}}(G+\ap {\Fh}+\ap^2 \Dh\phih_a \Dh\phih^a)}
+\frac{1}{4}C\Fh\widetilde{\Fh}\nonumber\\ 
& & +\,\, O(\pa\Fh,\pa \Dh\phih)~\label{NCDBIH}
\eeqr
Since (\ref{DBIH}) with $~\phi^a=g_s^{\frac{1}{4}}\phi^a_E~$ can also be 
obtained from four dimensional DBI with
metric $\mbox{g}_E=g_E+\pa{\phi_a}_E\pa\phi_E^a$  
as in (\ref{L+higgs}), we have that (\ref{DBIH}) is self-dual under the 
$SL(2,R)$ action  ${\phi^a_E}'=\phi_E$ and (\ref{Srot}).
We also define $\CL_\2$ as in (\ref{DBIr}) provided that each time 
we set $F=0$ we also set $\phi^a=0\,$.  $\CL_\2$ differs from (\ref{DBIH})
just by a constant term and a total derivative term. 
Formula (\ref{5}) is then obtained from 
the self-duality condition for $\CL_\DBI(F+B,\phi,g,g_s,C)$ discarding
the terms that do not contain $\phi^a$ and have order $0$ or $1$ in $F$.
As in (\ref{6}) we conclude that $\CL_\2(F,\phi_E,g_E,B,S)$ 
is self-dual w.r.t. $F$.
We now rewrite $\CL_\2$ as $\CL_\2(F,\phi,G,\th,G_s,C)$;
for $\theta$ light-like $\CL_\2(F,\phi,G,\th,G_s,C)$ is self-dual 
under  ${\phi^a_E}'=\phi^a_E$ and (\ref{GETrot}) 
so that (\ref{NCDBIH}) is self-dual.
\sk
In the $\ap\rightarrow 0$ limit, (\ref{NCDBIH}) gives noncommutative
electromagnetism coupled to Higgs fields
\beqr
\widehat{\CL}
&\equiv&\frac{\sqrt{G}}{G_s}\;\left( \frac{-1\,}{4}{\Fh}^2 
-\frac{1}{2} \Dh\phih_a\Dh\phih^a\right)
 \, +\,\frac{1}{4}C 
F\tilde{F}\\[1.1em]
&=&\CL_{\ap\!\!\rightarrow 0}+O(\pa F)+\rm{tot. ~der.}\nonumber
\eeqr
where the equality in the last line is the $\alpha\rightarrow 0$ limit of 
(\ref{Comm;NCDBIH}), and where now $\CL_{\ap\!\!\rightarrow 0}$ with $\th$ 
light-like and $1-\frac{1}{2}\th F>0$ is given by
\beqr
\CL_{\ap\!\!\rightarrow 0}&=&
\frac{\sqrt{G}}{G_s}\;
\frac{-\frac{1}{4}F^2-
\frac{1}{8}\th F^*\,FF^* -\frac{1}{2}{\rm{tr}}(\pa\phi_a\th FF\th\pa\phi^a)}{1-
\frac{1}{2}\th F}\nonumber\\[0.8em]
& &+\,\frac{\sqrt{G}}{G_s}\,\left[
{\rm{tr}}(\pa\phi_a F\th\pa\phi^a)-\frac{1}{2}(1-
\frac{1}{2}\th F)\,\pa\phi_a\pa\phi^a\right]\, +\,\frac{1}{4}C F\tilde{F}\label{LH}~.
\eeqr
Compatibility of the decoupling limit with duality rotations implies
that this Lagrangian  is self-dual under  
${\phi^a_E}'=\phi^a_E$ and (\ref{GETrot}). We conclude that (in the  
slowly varying field approximation) NCEM with scalar fields is self-dual.
\sk
We end this section observing that the change 
$\th\rightarrow \th'$ under a duality rotation, can be cancelled 
by a rotation in space and a $\th$ rescaling, or also by a Lorentz 
rotation in space-time.

\def\th{\theta}
\section{Gauge theory with space-time\\ noncommutativity}
We have seen that self-duality of NCDBI and NCEM requires light-like
noncommutativity. In this section we study a symmetry of static 
solutions of noncommutative Lagrangians, we call it $\th$-$\thE$
symmetry. To any space-noncommutative solution we associate
a space-time, and in particular light-like, noncommutative solution
and viceversa. In NCDBI and NCEM noncommutative self-duality  acts 
also within space-noncommutativity, it is a duality rotation 
obtained first going to light-like noncommutativity, then applying 
a duality rotation and finally going back to space noncommutativity.
In Subsection 3.2 and 3.3 we study how the $\th$-$\thE$ 
symmetry of NCDBI acts on the commutative DBI theory. New BPS type
solutions of DBI with both electric and magnetic backgrounds,
and their corresponding noncommutative solutions, are presented in
Subsection 3.5. 
\sk

We use the following notations:
$\Th$ is a generic constant noncommutativity tensor, we have 
$[x^\mu\*_{\!\!\!{\textstyle{,}}}\,x^\nu]=i\Th^{\mu\nu}\,$;
$\th$ is just a space-noncommutativity tensor $\th^{ij}$, 
$\th^{0i}=0\,$;
$\thE$ is a space-time noncommutativity tensor obtained from $\th$
adding electric components, $\thE^{ij}=\th^{ij}$, $\thE^{0i}$.
In three vector notation the electric and magnetic components of $\th$,
respectively $\thE$, are $(0,\thth)$ and $(\EE,\thth)$.
The background fields corresponding to $\Th,\th,\thE$ are 
$B,b,\bE$. We also recall that we have set $2\pi \ap=\ap\,,\;2\pi g_s=g_s$.
\sk

\subsection{Static solutions of space-time noncommutative gauge theory 
and SW map}

Consider a noncommutative Lagrangian $\CLh^\th=\CLh(\Fh,\phih,G,\star_\th)$.
The EOM for $\CLh^\th$ read 
\eq
f_\al(\Fh,\phih,G,\star_\th)=0 \label{space}
\en
where $f_\al$ are functions of the noncommutative fields and their derivatives.
We notice that a static solution of the $\CLh^\th$ EOM (\ref{space}) is also a static solution of 
(\ref{space}) with $\thE$ instead of $\th$, i.e. it is a static solution of the  $\CLh^\thE$ EOM. Indeed the star products 
$\star_\th$ and
$\star_\thE$ act in the same way on time independent fields.
Also the energy and charges of the solution are invariant.
A similar symmetry property holds if the fields are independent from a coordinate $x^\mu$ (not necessarily $t$).
This $\th$-$\thE$ symmetry property of static solutions can be used to construct moving solutions of a space-noncommutative theory $\CLh^\th$ from static solutions of the same theory
$\CLh^\th$.  
Indeed, given  $\th$,  if we turn on an electric component such that 
${\EE}\perp \thth$ and $|\EE | < | {\thth}|$, 
then with a Lorentz boost we can transform $\thE$ 
into a  
space-like $\th '$ proportional to the initial $\th$. Rescaling 
$\th'\rightarrow \th$ we thus obtain 
a solution (moving with constant velocity) of the space-noncommutative 
Lagrangian $\CLh^\th$.\footnote{For example, 
without loss of generality, 
consider $G=\eta$ and $\th$ given by ${\th}^{12}=-\th^{21}\not=0$ 
with
all others ${\th}^{\mu \nu}$  
vanishing. Let also $\thE$ be given by 
${\thE}^{\,12}=\th^{12}, {\thE}^{\,02}\not=0$ 
all others ${\thE}^{\,\mu \nu}$ 
vanishing. The static solution $\phih(x;\th)$, $\Ah_{\mu}(x;\th)$
of $\CLh^\th$ is also a solution 
of $\CLh^{\thE}$. Under the Lorentz boost 
$x^\mu\rightarrow {x'}^\mu=\La^{\mu}_{\;\nu}x^\nu$ given 
by $t'=\ga (t+vx)\,,~x'=\ga (x+vt)\,,~y'=y\,,~z'=z$ 
with $v=-\thE^{02}/\th^{12}$ the only nonvanishing component of
${\thE}'$ is $({\thE}')^{12}=\th^{12}/\ga$. 
Now notice that for generic $\thE$ and Lorentz transformation $\La$
we have: If $\phih(x;{\thE})$ is a solution of $\CLh^{\thE}$ 
then 
$\phih'(x;\La\thE\La^T)\equiv
\phih(\La^{\,-1}x\/;\/\thE)$ is a solution of 
$\CLh^{\La\thE\!\La^T}$. Similarly for $\Ah_\mu$. 
In our case we obtain that
${\phih'}(t,x,y,z,\th/\ga)\equiv\phih(\ga(x-vt),y,z,\th)$
and $\Ah'_\mu$ solve $\CLh^{\th/\ga}$. 
Rescaling $\th\rightarrow \ga\/\th$ we conclude
that $\phih(\ga(x-vt),y,z,\ga\/\th)$ and 
$\La^{\nu}_{\;\mu}\Ah_\nu(\ga(x-vt),y,z,\ga\/\th)$ 
solve $\CLh^{\th}$.
Here $\CLh^{\th}$ is arbitrary, in particular we obtain the moving
solitons studied in \cite{Bak:2000ym},\cite{Lechtenfeld:2001uq}. 
For multi-solitons with arbitrary relative motion see 
\cite{Bak:2001im},\cite{Lechtenfeld:2001uq}.}
It is this symmetry property that underlies 
the family of moving solutions found in \cite{Bak:2000ym}. 
\sk
We now use SW map and study how the $\th$-$\thE$ 
symmetry acts in the commutative 
theory. We have to consider the two SW maps $\SW^\th$ and $\SW^\thE$. 
In general a static solution 
$\phih$, $\Ah_\mu$ is mapped by $\SW^\th$ and $\SW^\thE$ into two 
different commutative solutions, however if $\Ah_0=0$ then 
$\SW^\th=\SW^\thE$. This can be seen from the index structure of SW map.
In general we have [cf. paragraph before 
(\ref{DBIH}-\ref{NCDBIH})]
\beqr
A_\mu &=& \Ah_\mu+\sum_{n>s} 
\Th^{(n)}\pa^{(n+s)}\Ah^{(n-s)}\Ah_\mu\nonumber\\[-.6em]
& &\label{SWmap}\\[-.47em]
\phi&=&\phih  +\sum_{n>s} \Th^{(n)}\pa^{(n+s)}\Ah^{(n-s)}\phih\nonumber
\eeqr
where the number of times $n,\;n+s,\;n-s$ that $\Th,\;\pa,\;\Ah$
appear is dictated by dimensional analysis. In (\ref{SWmap}) 
we do not specify which $\pa$ acts on which $\Ah$ and we do not
specify the coefficients of each addend. Because of the index structure
we notice that $\Th^{0i}$ never enters (\ref{SWmap}) 
if $\phih,~\Ah$
are time independent and $\Ah_0=0$. 
The commutative fields $\phi,\;A_i$ corresponding
to $\phih,~\Ah_i$ ($i\not=0$) are solution of both 
$\CL^{\th}$ and $\CL^{\thE}$. Here $\CL^{\th}$ and $\CL^{\thE}$ are 
the commutative Lagrangians
corresponding to $\CLh^\th$ and $\CLh^\thE$ via SW map:
$\CL^\th\equiv\CLh[\Ah(A,\phi),\phih(A,\phi),G,\star_\th]+
\rm{tot.der.}$ (similarly for  $\CL^{\thE}$).
The results of this subsection holds also when $\th$ is 
$x^i$-dependent.
\sk
\subsection{Dirac-Born-Infeld Lagrangian}

We now apply this construction to the case of  Dirac-Born-Infeld 
Lagrangian that (for slowly varying fields) is invariant in form under 
SW map and the change of variables (\ref{Gsgs}). 
We thus conclude that the static fields $\phih,~\Ah_i$
solve both the $\CL_\DBI(F+b,\phi,g,g_s)$ and  the
$\CL_\DBI(F+\bE,\phi,\gE,\gEs)$ EOM. Actually we are free
to impose the invariance of the closed string coupling constant $g_s$  
under this $\th$-$\thE$ transformation. 
{}From (\ref{Gsgs}) we see that then the
open string coupling constant $G_s$ rescales into $G^{\varepsilon}_s$ where
\eq
G^{\varepsilon}_s=G_s\sqrt{\frac{\det{(G^{-1}+\thE/\al)}}{\det{(G^{-1}+\th/\al)}}}
\label{detgG}~~.
\en
\def\gEs{g_s}
One can be concerned that this $\th$-$\thE$ symmetry might hold only in the 
slowly varying fields approximation.
In this subsection we further study this symmetry, we will
see that it is an exact symmetry of  static configurations of 
DBI Lagrangian.
In order to have a more explicit formulation of the $\th$-$\thE$ symmetry
we consider (rigid) coordinate transformations  $x\rightarrow x''$ and $
x\rightarrow x'$ that respectively orthonormalize $\gE$ and $g$,
while preserving time independence of the transformed $\phi\,,\,A$ 
fields.
From now on we set the noncommutative 
open string metric $G=\eta$. 
We study the case where the electric and magnetic components 
of $\th^{\mu\nu}$ are perpendicular  $\EE\perp {\thth}$. 
(We could have chosen a more general case, the final result (\ref{Newsol}) is 
independent from the $\EE\perp {\thth}$ choice).
With ${\EE}\perp {\thth}$, without loss of generality we 
consider 
\eq
\th\equiv -\thE^{12}=\thE^{21}~~,~~~
\Epsi\equiv\thE^{02}=-\thE^{20}~~~~~\mbox{all others $~\thE^{\mu\nu}=0\,$}~.
\label{thep}
\en
Then 
\eq
\begin{array}{c}
\gE_{00}=-1-\frac{\e^2}{\ap^2+\th^2-\e^2}~,~~
\gE_{01}=\frac{-\e\th}{\ap^2+\th^2-\e^2}~,~~
\gE_{11}=1-\frac{\th^2}{\ap^2+\th^2-\e^2}~,~~\\
\gE_{22}=\frac{\ap^2}{\ap^2+\th^2-\e^2}~,~~
\gE_{33}=1~~;~~~~
\bE_{02}= \frac{\e}{\ap^2+\th^2-\e^2}~,~~
\bE_{12}= \frac{\th}{\ap^2+\th^2-\e^2}~~
\end{array}\label{gE}
\en
all other $\gE_{\mu\nu}\,,\;\bE_{\mu\nu}$ components vanishing.
Under 
\eq  
x\rightarrow x''=(\GaE)^{-1}x~;~~~~
\GaE\equiv
\left(
\begin{array}{cccc}
\sqrt{1-\frac{\e^2}{\ap^2 +\th^2}} & -\frac{\e\th}{\ap\sqrt{\ap^2+\th^2}}
    &  0   & 0           \\[.57 em]
  0 &  \sqrt{1+\frac{\th^2}{\ap^2}} &  0   & 0          \\
  0 &  0  &  \sqrt{1+\frac{\th^2-\e^2}{\ap^2}}   &   0        \\
  0 &  0  &  0   &  1         
\end{array}\label{Ga}
\right)
\en
[obtained composing (\ref{M}) with a Lorentz transformation]
we have
\eq
\gE\rightarrow \eta~\,,~~~\bE\rightarrow b''=\GaE^T\bE\GaE~
\en
\eq
\phi(x)\rightarrow\phi''(x'')=\phi(\GaE x'')~\,,~~~
A(x)\rightarrow A''(x'')=\GaE^T A(\GaE x'')~\label{phiApp}
\en
In (\ref{phiApp}) $\phi''\,,\,A''$ are ${x''}^0$ independent and 
$A''_0=0$ if $\phi\,,\,A$ are $x^0$ independent and $A_0=0$.
The nonzero components of 
$b''$ are
\eq
\begin{array}{c}
e''\equiv -{b''}_{02}= \frac{-\e}{\ap\sqrt{\ap^2+\th^2}}~~,~~~
b''\equiv b''_{12}={\frac{\th}{\ap^2}}\sqrt{1-\frac{\e^2}{\ap^2+\th^2}}~.
\label{defepp}
\end{array}
\en
Obviously the fields $\phi'',\,A''$ satisfy the
$\CL_\DBI(F''+b'',\phi'',\eta,\gEs)$  EOM  if $\phi,\,A$ satisfy the
$\CL_\DBI(F+\bE,\phi,\gE,\gEs)$  EOM.

In order to obtain the $x\rightarrow x'$ transformation we consider 
$\E=0$ in (\ref{gE})-(\ref{defepp}), indeed
\eq
\th=\left.\thE\right|_{\varepsilon=0}~~,~~~ 
g=\left.\gE_{}\right|_{\varepsilon=0}~~,~~~ 
b=\left.\bE_{}\right|_{\varepsilon=0}
\label{normal}
\en 
and under
\eq
x\rightarrow x'=\Ga^{-1}x\label{xtoxp}~~~,~~~~~
\Ga\equiv\left.\GaE_{}\right|_{\varepsilon=0}
\en
we have $g\rightarrow \eta$, $b\rightarrow b'=\Ga^T b \Ga$.
The transformed solution $\phi'\,,\,A'$ satisfies the EOM of 
$\CL_\DBI(F'+b',\phi',\eta,g_s)$. 
The nonzero component of $b'_{\mu\nu}$ is${^{\,}}$: $\,b'\equiv b'_{12}=\th/\ap^2$. 
\sk
We now compose the $\Ga^{-1}$ and $\GaE$ transformations
and conclude that given a static solution $\phi'(x')\,,\,A'(x')$
of $\CL_\DBI(F'+b',\phi',\eta,g_s)$ with $A'_0=0$ and ${b'}_{12}=b'$
(all others ${b'}_{\mu\nu}$ vanishing) then $\phi''(x'')\,,\,A''(x'')$
is a static solution of 
$\CL_\DBI(F''+b'',\phi'',\eta,\gEs)$ where $A''_0=0$ and
\eq
\U\equiv\Ga^{-1}\GaE~\,,~~~x''=\U^{-1}x'~,~~~~~
\en
\eq
\phi''(x'')=\phi'(\U x'')~\,,~~~A''(x'')=\U^T A'(\U x'')~,\label{newsol}
\en
the nonzero components of $b''_{\mu\nu}$, expressed in terms of $e''$ and of $b'$
are 
\eq
\,e''\equiv-{b''}_{02}~~,~~~b''\equiv{b''}_{12}=\sqrt{1-{\ap^2 e''^2}}\; b'~~.
\label{eppbp}
\en
The explicit expression of $\U$ is 
\eq
\U=
\left(
\begin{array}{cccc}
\sqrt{1-\ap^2{e''}^2} & {}^{\,}\ap^2{e''}^{\,}b'
    &  0   & 0           \\
  0 &  1 &  0   & 0          \\
  0 &  0  &  \sqrt{1-\ap^2{e''}^2}   &   0        \\
  0 &  0  &  0   &  1         
\end{array}
\right)\label{U}~~.
\en
\sk
Let $\CA'$ be the gauge potential of $\CF'=F'+b'$, the
$\phi'\,,\,A'$ and $b'$ transformations then read 
\eq
\phi''(x'')=\phi'(\U x'')~\,,~~~\CA_i''(x'')=\U^j_{~i} \CA'_j(\U x'')~
\,,~~~\CA_0''(x'')=e''x''_2~~ .\label{Newsol}
\en
If $\phi'\,,\,\CA'_i$ is a static solution of $\CL_\DBI$ then 
$\phi''\,,\,\CA_{\mu}''$ is a new static solution. 
Notice now that in this section we have never used that the $F'_{ij}$ 
that appear in $\CF'=F'+b'$
vanish at spatial infinity; therefore (\ref{Newsol}) with 
$\U^j_{~i}$ as in (\ref{U}) holds for arbitrary static fields 
configurations $\phi'\,,\,\CA'$ with $\CA'_0=0$.  
\sk
{}For future reference we summarise the relation between
the coordinate systems $x,x',x''$ in the following commutative diagram
\sk
\begin{picture}(0,0)%
\includegraphics{DIAGRAM.pstex}%
\end{picture}%
\setlength{\unitlength}{4144sp}%
\begingroup\makeatletter\ifx\SetFigFont\undefined
\def\x#1#2#3#4#5#6#7\relax{\def\x{#1#2#3#4#5#6}}%
\expandafter\x\fmtname xxxxxx\relax \def\y{splain}%
\ifx\x\y   
\gdef\SetFigFont#1#2#3{%
  \ifnum #1<17\tiny\else \ifnum #1<20\small\else
  \ifnum #1<24\normalsize\else \ifnum #1<29\large\else
  \ifnum #1<34\Large\else \ifnum #1<41\LARGE\else
     \huge\fi\fi\fi\fi\fi\fi
  \csname #3\endcsname}%
\else
\gdef\SetFigFont#1#2#3{\begingroup
  \count@#1\relax \ifnum 25<\count@\count@25\fi
  \def\x{\endgroup\@setsize\SetFigFont{#2pt}}%
  \expandafter\x
    \csname \romannumeral\the\count@ pt\expandafter\endcsname
    \csname @\romannumeral\the\count@ pt\endcsname
  \csname #3\endcsname}%
\fi
\fi\endgroup
\begin{picture}(5112,2177)(896,-2563)
\put(1861,-2491){\makebox(0,0)[lb]{\smash{\SetFigFont{12}{14.4}{rm}{\color[rgb]{0,0,0}$(x',\eta,b')$}%
}}}
\put(4005,-2311){\makebox(0,0)[lb]{\smash{\SetFigFont{12}{14.4}{rm}{\color[rgb]{0,0,0}$\U$}%
}}}
\put(5512,-2505){\makebox(0,0)[lb]{\smash{\SetFigFont{12}{14.4}{rm}{\color[rgb]{0,0,0}$(x'',\eta,b'')$}%
}}}
\put(5491,-669){\makebox(0,0)[lb]{\smash{\SetFigFont{12}{14.4}{rm}{\color[rgb]{0,0,0}$(x,\gE,\bE)$}%
}}}
\put(3976,-518){\makebox(0,0)[lb]{\smash{\SetFigFont{12}{14.4}{rm}{\color[rgb]{0,0,0}$\th$-$\thE$}%
}}}
\put(6008,-1531){\makebox(0,0)[lb]{\smash{\SetFigFont{12}{14.4}{rm}{\color[rgb]{0,0,0}$\GaE$}%
}}}
\put(1907,-691){\makebox(0,0)[lb]{\smash{\SetFigFont{12}{14.4}{rm}{\color[rgb]{0,0,0}$(x,g,b)$}%
}}}
\put(1869,-1523){\makebox(0,0)[lb]{\smash{\SetFigFont{12}{14.4}{rm}{\color[rgb]{0,0,0}$\Ga$}%
}}}
\put(896,-1508){\makebox(0,0)[lb]{\smash{\SetFigFont{12}{14.4}{rm}{\color[rgb]{0,0,0}${}\;{}$}%
}}}
\end{picture}
\vskip -3.4cm
\eq
\label{table}
\en
\sk\sk\sk\sk\sk\sk\sk
\noi The quantities on the left hand side can be 
obtained from those on the right hand side simply by setting $\E=0$.

\subsection{Symmetries in extended target space}
Using the $\th$-$\thE$ symmetry property of noncommutative 
Lagrangians we have constructed new solutions (\ref{Newsol}) 
of $\CL_\DBI$; the field strength in these solutions has a constant 
term $b''_{\mu\nu}$ with both electric and magnetic 
components. Here we illustrate the commutative $\CL_\DBI$ 
symmetry  underlying this family of solutions. As a consequence
we see that the $\th$-$\thE$ symmetry of $\CL_\DBI$
is exact and not restricted to the slowly varying fields approximation.
We split $\CF'$ in its electric and magnetic 
components 
$\CE',\CB'$.
We then consider the Legendre transformation 
of $\CL_\DBI$ with  $C=0$ [cf. (\ref{LDBIWZ})]
\eq
\Hc(\vCE',\vCH',\phi')=\frac{1}{g_s}\vCB' \ccdot \vCH' + \CL_\DBI~~,~~~
{\mbox{where}}~~~~
\vCH'_i= -g_s\frac{\pa\CL_\DBI}{\pa \vCB'^i} ~~\label{leg}
\en
For time independent fields $\phi',\CA'_\mu$ we have
[$\ap=1$ in (\ref{LLL}) and (\ref{exp})]
\eq
\CL=\frac{-1}{g_s}
{\left[1+({\nabla}'\phi')^2+\vCB'^2-\vCE'^2+
({\nabla}'\phi'\ccdot\vCB')^2+
({\nabla}'\phi'\ccdot\vCE')^2
-({\nabla}'\phi')^2 \vCE'^2-
(\vCB'{\!}\ccdot\vCE')^2\right]^{\frac{1}{2}}}\label{LLL}
\en  
\eq
\CH'_i=\frac{\CB'_i+(\vCB'\ccdot{\nabla}'\phi')\,\pa'_i\phi'-
(\vCB'\ccdot\vCE')\,
\CE'_i}{\left[1+({\nabla}'\phi')^2+\vCB'^2-\vCE'^2+
({\nabla}'\phi'\ccdot\vCB')^2+
({\nabla}'\phi'\ccdot\vCE')^2
-({\nabla}'\phi')^2 \vCE'^2-
(\vCB'\ccdot\vCE')^2\right]_{{}_{{}_{}}}^{\frac{1}{2}}}\label{exp}
\en
We also have that  the EOM (\ref{3}) imply
$\vCH'=-{\nabla'}\chi'$
and  $\vCE'=-{\nabla'}\psi'$, ($\psi'=-\CA'_0$). 
As shown in  \cite{Gibbons:1998xz} it follows that  $\Hc(\psi',\chi',\phi')$
is the action  of a space-like 3-brane immersed in a 
target space of coordinates $X'^A=\{\ap\psi',\ap\chi',\ap\phi',{x'}^{\,i}\}$ 
and metric $\eta={diag}(-1,-1,1,1,1,1)$
\eq
\Sigma_{\Hc}\equiv\int\!\!d^3x' ~\Hc~~=\;~\frac{-1}{g_s\ap^2}\int\!\!d^3x'~ 
\sqrt{\det 
\left(\eta_{AB}\frac{\pa X'^A}{\pa x'^{\,i}}\frac{\pa X'^B}{\pa x'^{\,j}}\right)\,}
\label{H}~~.
\en
It is the $SO(2,4)$  symmetry \cite{Gibbons:1998xz} of this static gauge action 
that is relevant in our context: Consider the Lorentz transformation 
$Y^A=\Lambda^A_{~B}X'^B$ (where $X'^{\,i}=x'^{\,i}$) and express $Y^A$ as 
$Y^A=Y^A(y^i)$  (where $Y^i=y^i$) so that we are still in static gauge;
the action (\ref{H}) is invariant under $X'^A(x'^{\,i})\rightarrow Y^A(y^i)$.
In particular  consider a boost in the $\ap\psi'\,,\;x'_2$ 
plane with velocity $\beta=- \ap e''$. Under this boost
\eq
\begin{array}{c}
\left(\!
\matc
\ap\psi' \\
x'_2
\emat
\!\right)\;
\longrightarrow\;
\left(\!
\matc
\ap\psi'' \\
x''_2
\emat
\!\right)
=
\left(
\matcc
\ga &    \ga\be\\
\ga\be & \ga  
\emat
\right)
\left(
\!\matc
\ap\psi'\\
x'_2
\emat \!
\right)~\\[1.3em]
x''_1=x'_1~,~~x''_3=x'_3~~,~~~\chi''(x'')
=\chi'(x')~~,~~~\phi''(x'')=\phi'(x') 
\end{array}
\label{zero}
\en
where $\ga^{-1}=\sqrt{1-\be^2}=\sqrt{1-{\ap}^2e''^2}$.  
If $\psi'=0$ we have
\eq
x'^{\,i}=\U^i_{~j}x''^{\,j}
~~,~~~\psi''=- e'' x''_2\label{und}
\en
\eq
{}~~~~\CH''_2(x'')=\ga^{-1}\CH'_2(x')~~,~~~\CH''_\q(x'')=\CH'_\q(x')
{}~~~~~~~~~\sma{$\q=1,3$}\label{uno}
\en
\eq
~~\,{\pa}''_2\phi''(x'')=
\ga^{-1}\pa'_2\phi'(x')~~,~~~
\pa''_\q\phi''(x'')=\pa'_\q\phi'(x')\;~~~~~\sma{$\q=1,3$}\label{due}~~.
\en
We want to show that the magnetic induction 
\eq
{\vCB''^{{}^{\,}i}}=g_s\,\frac{\pa \Hc(\CE'',\CH'',\phi'')}{\pa \CH''_i}
\label{defBofH}
\en
 is the same as that 
obtained from (\ref{Newsol}), i.e.
\eq
{}~~~~~~~~~~\CB''_2(x'')=\CB'_2(x')~~,~~~\CB''_\q(x'')=\ga^{-1}\CB'_\q(x')~~~~~~\sma{$\q=1,3$}\label{tre}
\en
We can equivalently show that (\ref{due}) and (\ref{tre}) imply 
(\ref{uno}). This is done  using the explict expression (\ref{exp}).
\sk
We end this subsection observing that the $SO(2)$ 
subgroup of $SO(2,4)$ that acts on $\psi',\chi'$ 
is simply the 
$SL(2,R)$ subgroup of duality rotations (\ref{F+Brot})
that leaves invariant $g_s$ and $C=0$.

\subsection{BPS solutions for (NC)DBI with a scalar field}
In order to study BPS solutions \cite{Bogomolny:1976de} we minimize the energy functional 
(Hamiltonian) relative to NCDBI theory. The  Hamiltonian formalism
in space-time noncommutative theories is studied in \cite{Gomis:2001gy}. 
For theories (like NCDBI) invariant under time translations
the Hamiltonian defined in \cite{Gomis:2001gy} 
is the generator of this symmetry.
Here we are interested in the static case with 
$\Ah_0=0$, then $\Dh_0\phih=0$, $\Eh_i=-\Fh_{0i}=0$ and 
the Hamiltonian becomes 
as usual just minus the integral over space of the Lagrangian density
\beqr
\mbox{\bf{$\Sigmah_\DBI$}}&\!\!\!=&\!\!\!\int\!\! d^3x\,-\!\CLh_\DBI \,=\,
\int \!\!d^3x\,\frac{1}{\ap^2 G_s}\sqrt{{\rm{det}}(\eta_{ij}+
\ap {\Fh_{ij}}+\ap^2 \Dh_i\phih 
\Dh_j\phih)}+ O(\pa\Fh,\pa \Dh\phih)\nonumber\\
&\!\!\!=&\!\!\!\int\!\! d^3x\, \frac{1}{\ap^2 G_s}\sqrt{1+\ap^2(\DDh\phih)^2+\ap^2\BBh{}^2+
\ap^4(\BBh\ccdot\DDh\phih)^2}\,+ O(\pa\Fh,\pa \Dh\phih)\label{NCH}
\eeqr
Notice that $-\ap^2G_s\CLh\geq 1+\ap^2\left|\BBh\ccdot \DDh\phih\right|$,
so that 
\eq
\mbox{\bf{$\Sigmah_\DBI$}}\geq  |\hat{Z}_m|+ \int{\!\!d^3x~\frac{1}{\ap^2 G_s}}
\en 
where $\hat{Z}_m$ is the topological charge given by 
\eq
\hat{Z}_m=  \frac{1}{\ap^2 G_s} \int{\!\!d^3x\,\BBh\ccdot  
\DDh\phih}=  \frac{1}{\ap^2 G_s} \int{\!\!d^3x\,\DDh\ccdot(\BBh\phih)}~;
\en 
in the last equality we have used the Bianchi 
identity $\DDh\ccdot\BBh=0$. 
We rewrite the argument of the square root as 
$(1\pm \ap^2\BBh\ccdot\DDh\phih)^2+\ap^2(\BBh\mp\DDh\phih)^2$,
for fixed $\hat{Z}_m$ we see that 
$\mbox{\bf{$\Sigmah_\DBI$}}$ is minimal if the BPS equation 
\eq
\Bh_i=\pm\Dh_i\phih \label{NCBPS}
\en 
is satisfied (with plus  sign if $\hat{Z}_m>0$, minus sign otherwise).
The energy of this configuration, apart from the addend 
$(\ap^2G_s)^{-1}$ (D3-brane tension), 
equals the absolute value of the topological charge. 

Equations (\ref{NCBPS}) are also the BPS equations for noncommutative 
electromagnetism whose energy functional is
\beqr
\Sigmah_\EM&=&\frac{1}{2 G_s}\int\!\!d^3x\; \BBh\star\BBh+\DDh\phih\star\DDh\phih\\
&=&\frac{1}{2 G_s}\int\!\!d^3x\; (\Dh_i\phih\pm \Bh_i)\star (\Dh^i\phih\pm 
\Bh^i)
\,\mp\,\pa_i(\Bh^i\star\phih+\phih\star\Bh^i)\nonumber
\eeqr
The BPS equations for the commutative DBI Lagrangian 
$\CL_\DBI(\CF,\phi,g,g_s)$ are 
\cite{Callan:1998kz,Gibbons:1998xz,Gauntlett:1998ss}
$\CE_i=-\CF_{0i}=0$ and
\eq
\CB_i=\pm \pa_i \phi \label{BPS}
\en
they can be obtained as in the noncommutative case.
Here we have considered a metric $g_{\mu\nu}$ with block diagonal form 
$g_{00}=-1\,,\;g_{0i}=0$, so that $\CB_i=g_{ij}\CB^j$ 
with $\CB^i=\CF^{*\,0i}=\frac{1}{2}\sqrt{g^{-1}}\epsilon^{ijk}\CF_{ij}$.
The Bianchi identity $\pa_iB^i=0$ implies $\pa^i\pa_i\phi=0$; DBI monopoles 
corresponds to isolated singularities of the harmonic function $\phi$.
{}Using the definition of $\CH$ [cf. (\ref{leg}) and (\ref{exp})]
we see that (\ref{BPS}) is equivalent to 
\eq
\CH_i=\pm\pa_i\phi~~,~~~\pa^i\pa_i\phi=0
\label{HequalB}\;~.
\en
The energy of a BPS state equals the charge $Z_m=\frac{1}{\ap^2g_s}\int\! 
\CB^i\pa_i\phi=\frac{\pm 1}{\ap^2g_s}\int\! \CB^i\CH_i\,.$

A  solution to (\ref{BPS}) 
in a reference system  with metric $g$ and background $b$
as in (\ref{table})
is
\eq
{\phi}=\frac{{\,}-\th{\;}}{\ap^2}\,{x}^3 - \frac{1}{2 r}~~,~~~
\CB_i=-\pa_i\phi~~;~~~
r^2=g_{ij}x^i x^j
=({x}^3)^2+
\frac{(x^1)^2+(x^2)^2}{1+\th^2/\ap^2} \label{bpssol}  
\en
where $\frac{1}{2 r}=\frac{q_m}{4\pi r}$ with $q_m$ the magneton 
charge $q_m=2\pi$.
This solution describes a D1-string ending on a D3-brane.
Because of the magnetic backgound field $b$ on the brane the 
string is not perpendicular to the brane, in (\ref{bpssol})
the string is vertical and the brane is tilted, see Fig.$\!$ 2
(with axes $\bar{x}^3$ and $\ap\bar{\phi}$ renamed 
$x^3$ and $\ap\phi$).   
We call ${\pi/2}\,-\zeta$ the angle between the string and 
the brane, $\tan\!\zeta=\th/\ap$. The magnetic force acting 
on the end of the D1-string is compensated by the tension of the
D1-string \cite{Hashimoto:1999xh}.  
\sk

A solution of (\ref{NCBPS}) with  space-like noncommutativity
given by $\th^{12}=-\th$, (all others $\th^{\mu\nu}=0$) has been 
studied in \cite{Gross:2000wc}. It is a static solution with $\Ah_0=\Ah_3=0$.
Far away from the origin
the $\Bh$ field is the sum of two terms. 
The first term is that of a point-like negative charged magneton
placed at the origin, the second term vanishes for 
$x_3<<1$ while for $x_3>>1$ we have
\eq
\Bh_3=\frac{{\,}2{\,}}{\th}\,e^{-\frac{(x^1)^2+(x^2)^2 }{\th}}~~,~~~
\phih=\frac{-2x^3}{\th}\,e^{-\frac{(x^1)^2+(x^2)^2 }{\th}}~~
\label{fluxon}
\en
The energy in the interval $|x^3|\leq L$ with $L>>1$ grows
linearly with $L$ and is given by $\Sigmah=\frac{2\pi L}{G_s \th}$.
As we see from (\ref{fluxon}) this energy  is concentrated around 
the  positive $x^3$ axis.  We can think of a semi-infinite string  
along the positive $x^3$ axis with 
a tension
\eq
\hat{T}=\frac{2\pi}{G_s\,\th}\label{T}~~;
\en
attached to it there is a magnetic monopole smeared 
around the origin.
This configuration is shown to describe the D1-string 
ending on the  D3-brane  (\ref{bpssol}).
A first evidence is the correspondence between the tension of the
noncommutative string and that of the D1-string \cite{Gross:2000wc}, then 
in \cite{Gross:2000ph} the spectrum of small fluctuations around (a limit of) 
this solution is studied and found in agreement with the expectations 
from string theory. This suggests that the noncommutative BPS equations 
(\ref{NCBPS}) corresponds, via SW map, to the commutative BPS ones 
(\ref{BPS}).
More precisely \cite{Seiberg:1999vs,Moriyama:2000mm,Moriyama:2000eh},
for space noncommutativity 
the noncommutative BPS equations (\ref{NCBPS}) are
mapped into the nonlinear BPS equations of DBI theory. 
These nonlinear BPS equations are in turn
equivalent to the linear BPS equation (\ref{BPS}) via
a target space rotation in the $\ap\phi\,,\,x_3$ plane. 
If the magnetic part of $\th$ is given by $\th^{12}=-\th$ 
as in (\ref{bpssol}) the  $\ap\phi\,,\,x_3$ 
rotation is by an angle $\zeta$, $\tan\!\zeta=\th/\ap$
\beeq
\left(\!
\matc
\ap\phi \\
{x}^3
\emat
\!\right)\;
\longrightarrow\;
\left(\!
\matc
\ap\bar\phi \\
\bar{x}^3
\emat
\!\right)
=
\left(
\matcc
\cos\!\zeta &    -\sin\!\zeta\\
\sin\!\zeta & \cos\!\zeta  
\emat
\right)
\left(
\!\matc
\ap\phi\\
x^3
\emat \!
\right)~\label{zetarot}
\eeqn
we have \cite{Moriyama:2000mm}\footnote{In \cite{Moriyama:2000mm} the closed string 
variables are 
$\{g^M=\eta\;,b^M\}$ so that the open ones are 
$\{G^M\not=\eta\;,\th^M\}$. On the other hand 
we have chosen $\{G=\eta\;,\th\}$ so that $\{g\not=\eta\;,b\}$. 
The rotation angle $\zeta$ is the same $\tg\!\zeta=\th/\ap=\ap b^M$ 
in both descriptions because the coordinate transformation that maps 
$\{g^M,b^M\}\rightarrow \{g,b\}$ commutes with the $\zeta$-rotation.
}
\eq
\phih(x^1,x^2,x^3)
\stackrel{\sma{\rm{SW map}}}{\longleftarrow\!\!\!\longrightarrow}
\phi(x^1,x^2,x^3) 
\stackrel{\sma{\rm{$\zeta$-rot.}}}{\longleftarrow\!\!\!\longrightarrow}
\bar\phi(x^1,x^2,\bar{x}^3)\label{trot}
\en 
where $\phi(x^1,x^2,x^3)$ satisfies the nonlinear BPS equation and 
$\bar\phi(x^1,x^2,\bar{x}^3)$ the linear one. The profiles of $\phi$ and 
$\bar\phi$ are described in the following figures
\sk
\begin{picture}(0,0)%
\includegraphics{IMBUTI.pstex}%
\end{picture}%
\setlength{\unitlength}{4144sp}%
\begingroup\makeatletter\ifx\SetFigFont\undefined
\def\x#1#2#3#4#5#6#7\relax{\def\x{#1#2#3#4#5#6}}%
\expandafter\x\fmtname xxxxxx\relax \def\y{splain}%
\ifx\x\y   
\gdef\SetFigFont#1#2#3{%
  \ifnum #1<17\tiny\else \ifnum #1<20\small\else
  \ifnum #1<24\normalsize\else \ifnum #1<29\large\else
  \ifnum #1<34\Large\else \ifnum #1<41\LARGE\else
     \huge\fi\fi\fi\fi\fi\fi
  \csname #3\endcsname}%
\else
\gdef\SetFigFont#1#2#3{\begingroup
  \count@#1\relax \ifnum 25<\count@\count@25\fi
  \def\x{\endgroup\@setsize\SetFigFont{#2pt}}%
  \expandafter\x
    \csname \romannumeral\the\count@ pt\expandafter\endcsname
    \csname @\romannumeral\the\count@ pt\endcsname
  \csname #3\endcsname}%
\fi
\fi\endgroup
\begin{picture}(6391,2593)(17,-1643)
\put(1203,681){\makebox(0,0)[lb]{\smash{\SetFigFont{12}{14.4}{rm}{\color[rgb]{0,0,0}$\ap\phi$}%
}}}
\put(2704,-53){\makebox(0,0)[lb]{\smash{\SetFigFont{12}{14.4}{rm}{\color[rgb]{0,0,0}${x}^3$}%
}}}
\put(2073,-836){\makebox(0,0)[lb]{\smash{\SetFigFont{12}{14.4}{rm}{\color[rgb]{0,0,0}$\,\displaystyle\frac{\displaystyle\pi}{\displaystyle2}-\zeta$}%
}}}
\put(4507,-1584){\makebox(0,0)[lb]{\smash{\SetFigFont{12}{14.4}{rm}{\color[rgb]{0,0,0}\small{Fig. 2}}%
}}}
\put(6054,250){\makebox(0,0)[lb]{\smash{\SetFigFont{12}{14.4}{rm}{\color[rgb]{0,0,0}$\bar{x}^3$}%
}}}
\put(847,-1589){\makebox(0,0)[lb]{\smash{\SetFigFont{12}{14.4}{rm}{\color[rgb]{0,0,0}\small{Fig. 1}}%
}}}
\put(4854,794){\makebox(0,0)[lb]{\smash{\SetFigFont{12}{14.4}{rm}{\color[rgb]{0,0,0}$\ap\phi$}%
}}}
\put(4366,704){\makebox(0,0)[lb]{\smash{\SetFigFont{12}{14.4}{rm}{\color[rgb]{0,0,0}$\ap\bar\phi$}%
}}}
\end{picture}

As shown in \cite{Hashimoto:2000cf}, 
the $SO(2)$ rotation group (\ref{zetarot}) 
is a symmetry of DBI action. For static configurations 
the $SO(2)$ rotation group (\ref{zetarot})  
is a subgroup of the $SO(2,4)$ symmetry group 
discussed after formula (\ref{exp}). 
Actually the $\zeta$-rotation (\ref{zetarot}) commutes with 
the $\Ga$, $\U$ and $\GaE$ transformations, so that
[recall diagram (\ref{table})] it commutes with the $\th$-$\thE$
transformation: the $\zeta$ angle is independent from
the noncommutativity parameter $\E$. 
Since the rotation (\ref{zetarot}) is a symmetry
we have that the energy of a  field 
configuration is left invariant under (\ref{zetarot}).
It is instructive to see this explicitly for static field configurations,
some of the techniques in (\ref{zr})-(\ref{HamiltLeg}) will be used later.
We also show that the magnetic and electric charges  are independent from the 
$\zeta$-rotation.
Under the $\zeta$-rotation (\ref{zetarot}) we have 
\eq
\xb^1=x^1~,~~\xb^2=x^2~,~~\bar{\chi}({\xb})=\chi(x)~,~~\bar{\psi}(\xb)=
\psi(x)~,~~\bar{\pa_i}=\frac{\:\pa x^j}{\!{\pa}\xb^i}\,\pa_j\equiv J_i^{~j}\,\pa_j~\label{zr}
\en 
and recalling that $\CH_i=-\pa_i\chi$
and  $\CE_i=-\CF_{0i}=-\pa_i\psi$,
\eq
\bar{\CH}_i(\xb)=J_i^{~j}\CH_j(x)
~~,~~~\bar{\CE}_i(\xb)=J_i^{~j}\CE_j(x)~.
\en
In terms of $\CE$ and $\CH$ the magnetic induction $\CB^i=\CF^{*\,0i}$
is given by [cf. (\ref{leg})], 
$\CB^i(x)=g_s\frac{\!\!\!\!\!\del \SH}{\del \CH_i\!(x)}$.
The relation between $\CB(x)$ and the $\zeta$-rotated field $\bar{\CB}(\xb)$
is
\eq
\CB^i(x)=g_s\frac{\!\!\!\!\!\del \SH}{\del \CH_i(x)}=
g_s\int\!\!d^3\bar{y}\,
\frac{\del \SH}{\del \bar{\CH}_j(\yb)}\,
\frac{\del \bar\CH_j(\bar{y})}{\del {\CH}_i(x)}
=\bar{\CB}^j(\xb) J_j^{~i}\; \det{J^\1}\label{71}
\en
where in the second passage we have used that 
$\frac{\del \bar\CE}{\del {\CH}}=
\frac{\del \bar\phi}{\del {\CH}}=0$
because $J$ does not depend on $\CH$.
We then have
\eq
\pa_i\CB^i(x)\,d^3x=\pa_i\bar{\CB}^j(\bar{x})~J_j^{~i}\;{{}^{\!}}d^3\bar{x}
\,+ \, \bar{\CB}^j(\bar{x})\,\pa_i\!\left(J_j^{~i}\;\det{J^\1}\right) d^3x
=\bar{\pa_i}\bar{\CB}^i(\bar{x})\, d^3\xb\label{72}
\en
in the last passage we used $\pa_i{{}^{\,}}\det{J^\1}
=\pa_i({J^\1\!\!}_k^{~\l})\,\frac{\pa \det J^\1}{\pa {J^\1\!\!}_k^{~\l}}
=\pa_k({J^\1\!\!}_i^{~\l})~{J}_{\l}^{~k}\,\det{J^\1}\,$.
The same analysis holds if in (\ref{71}), (\ref{72}) we 
replace $\CH$ with $\CE$ and thus see that also the electric charge
is independent from the $\zeta$-rotation
\eq
\frac{1}{g_s}\pa_i\CD^i(x)\,d^3x=\frac{1}{g_s}\bar{\pa_i}\bar{\CD}^i(\bar{x})\, d^3\xb~,
\en 
here 
\eq
\CD^i=g_s\frac{\pa\CL_\DBI}{\pa\CE_i}=g_s\frac{\pa\Hc}{\pa\CE_i}~~.
\label{defD}
\en
Similarly we have
\beqr
\CB^i\CH_i\,d^3x&=&\bar{\CB}^i\bar{\CH_i}\, d^3\xb\label{BH}~~,\\
\CE_i\CD^i\,d^3x&=&\bar{\CE}_i\bar{\CD^i}\, d^3\xb\label{ED}~~.
\eeqr
Recalling (\ref{leg}) and the invariance of (\ref{H}) under 
the $\zeta$-rotation 
(\ref{zetarot}) we obtain that $\CL_\DBI d^3x$ is invariant. 
The invariance of the energy functional follows from
\eq
H(x)d^3x=\bar{H}(\bar{x})d^3\bar{x}\label{invH}
\en
where the energy density $H$ is given by 
the Legendre transformation
\eq
H=\frac{1}{g_s}\CE_i\CD^i-\CL_\DBI\label{HamiltLeg}~\;. 
\en
In (\ref{HamiltLeg}) $H$ equals the gravitational energy density $-T^0_{~0}$,
with
$T_{\mu\nu}=\frac{-2}{\sqrt{g}}\frac{\pa \CL_\DBI}{\pa g^{\mu\nu}}$ 
\cite{Born:1934gh}.
{}From (\ref{invH}) one can also show that the string tension
is independent from the $\zeta$-rotation.
\sk

Since the $\zeta$-rotation (\ref{zetarot}) 
is independent from the noncommutativity
parameter $\E$, and since it
does not modify the energy,
the electric and magnetic charges 
$\frac{1}{g_s}\int\!\!d^3\!x\,\pa_i\CD^i\,$,
$\,\int\!\!d^3\!x\sqrt{g}\,\pa_i\CB^i$ 
and the string tension of a given field configuration,
in the following we call SW map what is actually SW map plus  
$\zeta$-rotation; we also write 
$\phi,x^3$ instead of $\bar\phi,\bar{x}^3$. 

\subsection{BPS type solutions with both electric and magnetic 
backgrounds}
From Subsection 3.1 and (\ref{detgG})  we have that  \GN 
soliton is also a BPS solution of space-time noncommutative DBI
theory. The string tension is now
\eq
\hat{T}^\varepsilon=\frac{2\pi}{\GEs\,\th}\label{TE}~~.
\en
Applying SW map to this  noncommutative BPS solution we 
obtain that (\ref{bpssol}) with 
$\CE_2=-\bE_{02}$, $\CE_1=\CE_2=0$ solves 
$\CL_\DBI(F+\bE,\phi,\gE,\gEs)$. More generally 
$\CL_\DBI(F+\bE,\phi,\gE,\gEs)$ admits the first order
equations (\ref{BPS}) that using tensor notation read
\eq
\CF_{\mu\nu}\equiv F_{\mu\nu}+\bE_{\mu\nu}=-\sqrt{g}\,
\epsilon_{\mu\nu\!\rho 0}\,g^{\rho\sigma}\frac{\pa \phi}{\pa x^\rho}
+\bE_{\mu\nu}-b_{\mu\nu} ~~~~~\label{BPSE}
\en
In the orthonormal reference system $x''$, recall diagram (\ref{table}),
equations (\ref{BPSE}) read 
\beqr
\CB''_2&=&-\gamma\,\pa''_2\phi''\nonumber\\
\CB''_\q&=&-\gamma^{-1}\,\pa''_\q\phi''\label{BPSpp2}~~~~~~~~~~~~~~
\sma{$\q=1,3$}\\
\CE''_2&=&e''~~~,~~~~\CE''_\q=0~~\nonumber
\eeqr
with $\pa^i\CB''_i=0$ and $\ga^{-1}=\sqrt{1-\ap^2 e''^2}\,$.$\;$
In order to obtain (\ref{BPSpp2}) from (\ref{BPSE})
multiply (\ref{BPSE}) by $\GaE^T$ and by $\GaE$ and then express 
all the remaining tensors in terms of $e''$ and $b''$ 
[cf. (\ref{defepp}), (\ref{eppbp})]. 
Equations (\ref{BPSpp2}) are of BPS type, they are first order
equations in the gauge potential and 
the scalar field. They correspond to the BPS equations of noncommutative
gauge theory $(\ref{NCBPS})$. Because of the commutativity of
the diagram  (\ref{table}) they can also be obtained from the BPS equations
in the $x'$ reference system 
\eq
\CB'_i=-\pa'_i\phi'~~~,~~~~\CE_i'=0~\label{BPS'}
\en
using the boost (\ref{zero}) in the $\psi',\,x_2'$ plane. 
[Is is straighforward to explicitly  derive (\ref{BPSpp2})
from (\ref{BPS'}) using (\ref{und}),(\ref{due}),(\ref{tre})].  
Since (\ref{BPS'}) is equivalent to (\ref{HequalB}) (with primed variables)
then, recalling (\ref{uno}) and (\ref{due}), we have that (\ref{BPSpp2}) is 
equivalent to
\eq
\CH''_i=-\pa''_i\phi''~~,~~~~
({\pa''_1}^{\,2}+\ga^2{\pa''_2}^{\,2}+{\pa''_3}^{\,2})\phi''=0
~~,~~~~\CE''_2=e''~,~~\CE''_1=\CE''_3=0~~.~\label{BPSeq}
\en
The boost (\ref{zero}) that relates (\ref{BPS'})  to 
(\ref{BPSpp2})
is a symmetry of the action $\int \!\! d^3x\,\CL_\DBI$.
Indeed under (\ref{zero}), for a generic static configuration, the 
functional $\Sigma_\Hc$ [cf. (\ref{H})] 
is invariant and from (\ref{71}) where now 
$J_i^{~j}=\frac{\;\pa x'^{\,j}}{\pa x''^{\,i}\:}$ we have, as in 
(\ref{72}),(\ref{BH})
\eq
\pa'_i\CB'^{\,i}\,d^3x'=\pa''_i\CB''^{\,i}\,d^3x''\label{Bcharge}
\en
\eq
\CB'^{\,i}\CH'_{i}\,d^3x'=\CB''^{\,i}\CH''_i\,d^3x''
\en
recalling (\ref{leg}) we then have
\eq
\CL(\CE',\CB',\phi')\,d^3x'=\CL(\CE'',\CB'',\phi'')\,d^3x''~.\label{L''L'}
\en
Also the canonical energy $\int\!d^3x' H_c=\int\!d^3x'\, T_c^{00}\,$,
obtained from the canonical energy momentum tensor 
$T_c^{\mu\nu}=\frac{\pa \CL(\CA',\phi')}{\pa\,\pa_\mu\CA'_\rho}\,
\pa^\nu\!\CA'_\rho\,+ \eta^{\mu\nu}\CL$ is invariant under the boost 
(\ref{zero}), indeed for a generic static configuration we have $H_c=-\CL\,$.
As (\ref{Bcharge}) shows, the magnetic charge too is invariant under the boost 
(\ref{zero}).

Given a BPS state (\ref{BPS'}), the boosted one (\ref{BPSpp2}) has a charge
%
\eq
Z''_m=\int\!d^3x''\,\CB''^{\,i}\pa_i\phi''
=-\int\!d^3x''\,\CB''^{\,i}\CH''_i=-\int\!d^3x'\,\CB'^{\,i}\CH'_i
=\int\!d^3x'\,\CB'^{\,i}\pa_i\phi'=Z'_m
\en
and, recalling (\ref{HamiltLeg}), a gravitational energy 
\eq
\Sigma''=\int\!d^3x'' \,H(\CD'',\CB'',\phi'') \,=
\Sigma'+\frac{1}{g_s}\int\!d^3x''\,\CE''_i\CD''^{\,i}~.\label{93}
\en
We see that the energy is of BPS type, indeed it is the sum  of the old BPS
energy $\Sigma'$ plus the topological charge $Z''_e=\frac{1}{g_s}\int\!d^3x''
\,\CE''_i\CD''^{\,i}=-\frac{1}{g_s}\int\!d^3x''\,\pa_i\psi''\CD''^{\,i}\,$.
The explicit value of $\CD''^{\,i}$ [see definition (\ref{defD})] can be 
obtained
from (\ref{L''L'}),(\ref{due}),(\ref{tre}) expressing  $\CD''^{\,i}$
in terms of $\pa'_i\phi'$. We have 
\eq
\CD''^{\,2}= e''\gamma~~,~~~ \CD''^{\,1}=\CD''^{\,3}=0~.\label{ExplicitED}
\en
At second order in $e''$ the energy is 
$\Sigma''=\frac{1}{\ap^2g_s}\int\!\!d^3x''+|Z''_m|+\frac{{e''}^2}{2\, g_s}\int\!\!d^3x'' $,
we recognize the brane tension, the topological charge and the extra tension due to the energy of the electric field $e''$. The energy $\Sigma''$ to all orders in $e''$ is obtained just 
replacing in the last addend $\frac{1}{2}{e''}^2$ with the relativistic term $\ga-1$.
\sk
Solution (\ref{bpssol}) in the $x''$ reference system reads\footnote{If we let $\ga\rightarrow \ga^{-1}$ in (\ref{solpp}) 
 then $\CB''_i$ is the magnetic field of a  monopole (plus constant background $b'$) moving
with velocity $\beta$ in the $x''_2$ direction}
\beqr
\phi''&=&-\gamma\,b'' x''^{\,3}-\frac{1}{2 \, R}
\nonumber\\
\CB''^{\,i}&=&-\frac{1}{2 \,\gamma}\;\frac{x''^{\,i}}{R^3} + b''\del^{i3}
\label{solpp}\\
\CE''_2&=&e''~~~,~~~~\CE''_1=\CE''_3=0~~~,~~~~~~
R^2\equiv (x''^{\, 1})^{2}+
\gamma^{-2} (x''^{\, 2})^{2} + (x''^{\, 3})^{2}\nonumber
\eeqr
In the next subsubsection we show that the string tension associated to this 
configuration is  that of a D1-string, as we expect from a BPS state.
Notice that the shape of the funnel representing the string is
no more symmetric in the $x''_1,x''_2$ directions. A section 
determined by $\ap\phi''+\ap\ga b''x''_3=\,${\it const} is an ellipsis in the 
$x_2'',\rho$ plane, here $\rho=\sqrt{x_1''^2 +x_3''^2}_{{}_{{}_{}}}$. 
The ratio between the  ellipsis axes is given by
$\gamma=
\frac{1}{\sqrt{1-\ap^2{e''^2}}}=
\sqrt{\frac{\ap^2+\th^2}{\ap^2+\th^2-\e^2}}^{{}^{{}}}$

\begin{picture}(0,0)%
\includegraphics{OVALE.pstex}%
\end{picture}%
\setlength{\unitlength}{4144sp}%
\begingroup\makeatletter\ifx\SetFigFont\undefined
\def\x#1#2#3#4#5#6#7\relax{\def\x{#1#2#3#4#5#6}}%
\expandafter\x\fmtname xxxxxx\relax \def\y{splain}%
\ifx\x\y   
\gdef\SetFigFont#1#2#3{%
  \ifnum #1<17\tiny\else \ifnum #1<20\small\else
  \ifnum #1<24\normalsize\else \ifnum #1<29\large\else
  \ifnum #1<34\Large\else \ifnum #1<41\LARGE\else
     \huge\fi\fi\fi\fi\fi\fi
  \csname #3\endcsname}%
\else
\gdef\SetFigFont#1#2#3{\begingroup
  \count@#1\relax \ifnum 25<\count@\count@25\fi
  \def\x{\endgroup\@setsize\SetFigFont{#2pt}}%
  \expandafter\x
    \csname \romannumeral\the\count@ pt\expandafter\endcsname
    \csname @\romannumeral\the\count@ pt\endcsname
  \csname #3\endcsname}%
\fi
\fi\endgroup
\begin{picture}(3712,2575)(2336,-2585)
\put(3041,-1521){\makebox(0,0)[lb]{\smash{\SetFigFont{12}{14.4}{rm}$~~~~~~~~~~~~~$}}}
\put(2336,-1546){\makebox(0,0)[lb]{\smash{\SetFigFont{12}{14.4}{rm}$~~~~~$}}}
\put(5681,-1351){\makebox(0,0)[lb]{\smash{\SetFigFont{12}{14.4}{rm}$a$}}}
\put(5306,-511){\makebox(0,0)[lb]{\smash{\SetFigFont{12}{14.4}{rm}$\ga a$}}}
\put(5246,-166){\makebox(0,0)[lb]{\smash{\SetFigFont{12}{14.4}{rm}$x''_2$}}}
\put(5916,-1666){\makebox(0,0)[lb]{\smash{\SetFigFont{12}{14.4}{rm}$\rho=\sqrt{{x''_1}^2+{x''_3}^2}$}}}
\end{picture}
\nopagebreak
\sk
\vskip -4.15cm
\eq
\label{ellisse}
\en
\sk\sk\sk\sk\sk\sk
\noi
Since  $\ga b''=\th/\ap^2$, when $\ap\rightarrow 0$ 
with open string parameters held fixed  the brane-string 
angle goes to zero: the brane is now perpendicular to the  
$x''_3$ axis and the string lies on the brane.
The ellipsis (\ref{ellisse}) (now with $x''_3=0$) has the axis 
$a=\frac{-\ap}{2\; \mbox{\it{const}}}$ and therefore the string is eventually
localized at $x''_1=0$ for $\ap\rightarrow 0$ 
(this is the same as in the commutative case 
$\th=\E=0$). On the other hand, since for $\thE$ light-like
$\ga\sim\th/\ap$, 
 the other ellipsis axis is 
$\ga a=\frac{-\th}{2\; \mbox{\it{const}}}$ so that the string smoothly
opens into the brane along the $x''_2$ direction independently from 
$\ap\rightarrow 0$.

\subsubsection{String Tension}
In this subsection we match the noncommutative string tension
(\ref{TE})  with the string tension of configuration (\ref{solpp}) 
written in the $x$ reference system, i.e. of  (\ref{bpssol})
with $\CE_2=-\bE_{02}$, $\CE_1=\CE_2=0$. This last is a solution of the
$\CL_\DBI(F+\bE,\phi,\gE,\gEs)$ first 
order equations (\ref{BPSE}). 
We begin by reviewing the  string tension in the $x'$  
reference system, i.e. for configuration 
(\ref{solpp}) with $\E=0$.
Let $\ap\Phi'$ be the axis perpendicular to the brane, it has an
angle $\zeta$ with the string axis $\ap\phi'$; $\tg\!\zeta=\ap b'=\th/\ap$. 
We have (see Fig.~2, 
with axes $\bar{x}^3$, $\ap\phi$, $\ap\bar{\phi}$ renamed 
$x'_3$,  $\ap\Phi'$, $\ap\phi'$ ) 
\eq
\ap\Phi'=\cos\!\zeta\, \ap\!\phi'+\sin\!\zeta\,x'_3=-\ap\cos\!\zeta\,
\frac{1}{2 r'}
\en
where $r'^2={x'_1}^2+{x'_2}^2+{x'_3}^2$. The string tension $T'$ is
\eq
T'\,\frac{\Delta \ap\Phi'}{\cos\!\zeta}=\Delta\Sigma'\label{Tptension}
\en
where $\Delta\Sigma'=
\left.\Sigma'\right|_{\ap\!\Phi'}^{\ap\!\Phi'+\Delta\ap\!\Phi'}$.
We have
\beqr
\Delta_0\Sigma'&=&
\left.\Sigma'\right|_{\ap\!\Phi'_0}^{\ap\!\Phi'_0+\Delta\ap\!\Phi'_0}
=\frac{1}{g_s}\int_{V_0^{{}^{}}}\!\!\!\!d^3x'\,(\nabla'\phi')^2\nonumber\\
&=&\frac{1}{g_s\cos^2\!\zeta}\int_{V_0^{{}^{}}}\!\!\!\!d^3x'\,(\nabla'\Phi')^2 
+\frac{1}{g_s}\int_{V_0^{{}^{}}}\!\!\!\!d^3x'\, b'^2\nonumber\\
&=&\frac{1}{g_s\cos^2\!\zeta}\int_{\pa V_0^{{}^{}}}\!\!\!\!\!dS'\,
\Phi'\nabla'\Phi'+\frac{1}{g_s}V_0 b'^2\nonumber\\
&=&\frac{2\pi}{g_s\cos\!\zeta}\Delta\Phi'_0+\frac{1}{g_s}V_0 b'^2
\eeqr
where $V_0$ is the region in space such that $\Phi'$ is between $\Phi'_0$ and
$\Phi'_0+\Delta\Phi'_0$, and $\pa V_0$ is given by the sphere 
$\Phi'=\Phi'_0+\Delta\Phi'_0$ and the sphere $\Phi'=\Phi'_0$. 
In the fourth equality we integrated by parts and used that $\Phi'$ is 
divergence free in $V_0$. 
When $\Phi'_0>>1 $, then $V_0\sim 0$ if 
$\Delta\Phi'_0$ is fixed, it follows that the energy is simply proportional to 
$\Delta\Phi'_0$ and the string tension is that of a D1-string
\eq
T'=\frac{2\pi}{g_s \ap}
\en
In order to obtain the string tension associated to the field 
configuration (\ref{solpp}), we recall that $\phi''(x'')=\phi'(x')$ and
$x''_3=x'_3$, so that
$\Phi''(x'')=\Phi'(x')$. Then from 
(\ref{Tptension}),(\ref{93}),(\ref{ExplicitED}) we have
\eq
T''=T'=\frac{2\pi}{g_s \ap}\label{Tpptension}
\en
Similarly,
when $x''_3>>1$ then $\frac{1}{r'^4}\sim 0$, and we also have
\eq
\Delta W''\equiv\left.\Delta\Sigma''\right|_{x''_3}^{x''_3+\Delta x''_3}
=\left.\Delta\Sigma'\right|_{x'_3}^{x'_3+\Delta x'_3}\sim
\left.\frac{-1}{g_s}\int\!\!d^3x'\,\frac{b'x'_3}{r'^3}\,\right|_{x'_3}^{x'_3+\Delta x'_3}\!=
\frac{-2\pi}{g_s}\,b'\Delta x''_3\:
\en
$\Delta W''$  is a string-brane interaction energy stored
in between  $x''_3$ and $x''_3+\Delta x''_3$, it
is the work necessary to move the charge $-q_m=-2\pi$ from the point 
$\Cal{P}$ on the brane to the point ${\Cal{P}}+\Delta\Cal{P}$ on the brane, 
where the value of the third coordinate of 
$\Cal{P}$ and  ${\Cal{P}}+\Delta\Cal{P}$ is respectively $x''_3$ and 
$x''_3+\Delta x''_3$.
Since the brane for $x''_3>>1$ 
lies in the $\phi''+b' x_3''=0$ plane (recall $\ga b''=b'$), the charge $-q_m$ moves a distance 
$\frac{\Delta x''_3}{\cos\!\zeta}$, therefore to $\Delta W''$ is associated
a tension
\eq
t''=\frac{\Delta W''}{\Delta x''_3}\cos\!\zeta\label{tpp}
\en 
In the $x$ reference system, see (\ref{Ga}), we have, from the transformation
rule of the energy-momentum vector 
and from $x''_3=x_3$ and $\phi''(x'')=\phi(x)$
\eq
\Delta \Sigma=\gamma\,\Delta \Sigma''~~,~~~\Delta W=\gamma\,\Delta W''
\en
\eq
T=\ga T''~~,~~~t=\ga t''\label{Tt}
\en
As in \cite{Gross:2000wc} 
we project the energy of the string on the brane 
(we consider the shadow of the string on the brane) and obtain a total tension
along the string shadow
\eq
\frac{1}{\sin\!\zeta}T+t=\frac{2\pi \ap}{\sqrt{\ap^2+\th^2-\E^2}\,\th g_s}
=\frac{2\pi}{\GEs\th}\label{tensione}
\en
where in the last equality we used (\ref{Gsgs}) and (\ref{detgG}). The tension (\ref{tensione})
is the same as the noncommutative string tension (\ref{TE}) of \GN 
BPS soliton in the presence of $\thE$ noncommutativity. We thus confirm 
that SW map relates this $\thE$-noncommutative soliton to (\ref{solpp})
written in the $x$ reference system.

\subsubsection{Dual string-brane configuration}

We now duality rotate configuration (\ref{solpp}) 
and obtain a soliton solution that describes a fundamental 
string ending on a D3-brane with electric and magnetic background.
Under the rotation (\ref{F+Brot}) with
$a=d=0$,  $c=-b^{-1}=2\pi$ and $S_1=C=0$, $g_{\mu\nu}=\eta_{\mu\nu}$
we have 
\eq
({g_{YM}^{\d}})^{2}=g_s^{\d}=\frac{4\pi^2}{g_s}~~~,~~~~~C_\d=0 ~~~,~~~~~
g_{\mu\nu}^\d=\frac{2\pi}{g_s}\eta_{\mu\nu}~~~,~~~~~
\phi''_\d =\left(\frac{2\pi}{g_s}\right)^\frac{1}{2}\,\phi''~~
\label{phiDual}
\en
and from 
$\CE_i=-\CF_{0i}\,$, $\CB^i={\CF^{*}}^{0i}$, 
$\CD^i=-g_s\tilde{\CK}^{0i}$, $\CH_i=-g_s\sqrt{g} \CK_{0i}\,$ and 
(\ref{BPSpp2}),(\ref{BPSeq}),(\ref{ExplicitED}) we obtain the 
BPS type equations
\beqr
{\CD''_\d}^2&=&-\sqrt{g^\d}\left(\frac{2\pi}{g^\d_s}\right)^\frac{1}{2}
\ga\,\pa''^2\phi''_\d\nonumber\\
{\CD''_\d}^q&=&-\sqrt{g^\d}\left(\frac{2\pi}{g^\d_s}\right)^\frac{1}{2}
\ga^{-1}\pa''^q\phi''_\d\label{BPSDpp}~~~~~~~~~~~~~~
\sma{$\q=1,3$}\\[.5em]
{\CH''_i}^\d&=&\frac{g_s^{\d^{{}^{}}}}{2\pi}\sqrt{g^\d} e''\,\delta_{i2}
\eeqr
or equivalently
\eq
\CE''^\d_i=
-\left(\frac{g_s^{\d^{{}^{}}}}{2\pi}\right)^\frac{1}{2}\pa''_i\phi_\d
{}~\,,~~~~
({\pa''_1}^{\,2}+\ga^2{\pa''_2}^{\,2}+{\pa''_3}^{\,2})\phi''_\d=0
{}~\,,~~~~{\CB''_\d}^i
=-\frac{2\pi}{g_s^{\d^{{}^{}}}} \ga\,e''\delta^{i2}
{}~~.
\en
The dual of solution (\ref{solpp}) is given by (\ref{phiDual}) and 
\eq
\CA''^\d_0=-\left(\frac{2\pi}{g_s}\right)^\frac{1}{2}\phi''_\d~~,~~~
\CA''^\d_1=-\frac{2\pi}{g_s}\ga\,e''x''^3~~,~~~\CA''^\d_2=\CA''^\d_3=0
\label{ASWm1}~~~.
\en 

Consider now the $\thE$ light-like case. We then have $\det\GaE=1$ and 
duality rotations 
commute with the coordinate transformation $\GaE$. They also commute 
with the $\zeta$-rotation because this rotates $X=\ap\phi$ with $x^3$
and under duality $X$ and $x^3$ are unchanged [under duality 
$\ap\phi_\d\not=\ap\phi$ but $X_\d=X$ because it is the component 
$\eta_{XX}$ of the target space metric that changes according to 
(\ref{phiDual})].
In conclusion applying the $\GaE$ coordinate transformation,
the $\zeta$-rotation and SW map we obtain from (\ref{ASWm1}) and 
(\ref{phiDual}) the noncommutative fields $\Ah^\d,\phih_\d$ with 
light-like noncommutativity 
$\thE^{\mu\nu}_{\!\!\!\!\d}=\frac{g_s}{2\pi}\,\thE^{*\mu\nu}\,$. 
These noncommutative fields correspond to a fundamental string
ending on a D3-brane with light-like background.
The $\Ah^\d,\phih_\d$ fields solve also $\CLh_\DBI(\Fh,\phih,\*_{\th_\d})$
where $\th_\d$ is just the space part of $\thE_{\!\!\!\!\d}\,$,
they describe a space-noncommutative electric monopole with a string 
attached.

\sk
\sk
\noi{\bf{Acknowledgements}}\\
I wish to thank Sergei M. Kuzenko, Yaron Oz, Stefan Theisen, for 
helpful discussions. I acknowledge fruitful discussions with 
Branislav Jur\u{c}o, Olaf Lechtenfeld, Ivo Sachs and
Peter Schupp. I thank Leonardo Castellani for his  comments
and suggestions.
This work has been supported by Alexander von Humboldt-Stiftung.

\end{document}